\documentclass[reprint, amsmath, amssymb, superscriptaddress, prl]{revtex4-2}
\usepackage{graphicx}
\usepackage{hyperref}
\usepackage{xcolor}
\usepackage{bbm}

\begin{document}

\title{Quantum Optimal Control without Arbitrary Waveform Generators}
\author{Qi-Ming Chen}
\affiliation{QCD Labs, QTF Centre of Excellence, Department of Applied Physics, Aalto University, FI-00076, Espoo, Finland}
\affiliation{Department of Chemistry, Princeton University, Princeton, New Jersey 08544, USA}
\affiliation{Department of Automation, Tsinghua University, Beijing 100084, China}

\author{Herschel Rabitz}
\email{hrabitz@princeton.edu}
\affiliation{Department of Chemistry, Princeton University, Princeton, New Jersey 08544, USA}

\author{Re-Bing Wu}
\email{rbwu@tsinghua.edu.cn}
\affiliation{Department of Automation, Tsinghua University, Beijing 100084, China}

\date{\today}

\begin{abstract}
	Simple, precise, and robust control is demanded for operating a large quantum information processor. However, existing routes to high-fidelity quantum control rely heavily on arbitrary waveform generators that are difficult to scale up. Here, we show that arbitrary control of a quantum system can be achieved by simply turning on and off the control fields in a proper sequence. The switching instances can be designed by conventional quantum optimal control algorithms, while the required computational resources for matrix exponential can be substantially reduced. We demonstrate the flexibility and robustness of the resulting control protocol, and apply it to superconducting quantum circuits for illustration. We expect this proposal to be readily achievable with current semiconductor and superconductor technologies, which offers a significant step towards scalable quantum computing.
\end{abstract}

\maketitle

\emph{Introduction.}---The desire to manipulate and control quantum-mechanical phenomena has grown with the development of quantum information processors \cite{Warren1993, Rabitz2000}. Moving beyond the attempts at seeking physically intuitive control fields, quantum optimal control (QOC) opened the door to high-fidelity quantum operations with modulated waveforms \cite{Peirce1988, Kosloff1989, Judson1992, Assion1998}. For many years, QOC has been successfully applied to various physical platforms, for example, superconducting quantum circuits (SQC) \cite{Tian2000, Steffen2003, Motzoi2009, Safaei2009, Rebentrost2009, Schutjens2013, Zahedinejad2015, Chow2010, Chow2011, Chow2012, Heeres2017, Galiautdinov2012, Ghosh2013, Barends2014, Kelly2015, Roushan2017, Lucero2008, Kelly2014, Arute2019, Foxen2020, Harrigan2021}. Arbitrary waveform generators (AWGs) have become indispensable for high-fidelity quantum control, which are used routinely with SQC and some other quantum information realizations. However, towards the goal of building even larger quantum information processors, the brute-force scaling of current AWG technologies involves substantial overhead in experimental resources, and attempts at integrating arrays of AWGs at low temperature remains a technical challenge \cite{Hornibrook2015, Reilly2015, Vandersypen2017, Bardin2019, Dijk2019, Petit2020, Pauka2021, Xue2021}. Besides these efforts of pursuing the ability of generating more delicate waveforms, one may naturally ask whether QOC necessarily requires an arbitrary waveform? 

Here, we report that the control of a general quantum system can be realized by simply turning on and off the driving fields at prescribed time instances \cite{Chen2017}. This method is motivated by the pulse width modulation (PWM) technique in classical control systems, where the average power delivered to the plant is controlled by switching the supply at a fast rate and suitable pattern \cite{Holmes2003}. We generalize this idea to quantum-mechanical systems and reveal the equivalence between an arbitrary waveform and a PWM pulse train. We also show that a PWM sequence can be efficiently designed with existing QOC algorithms, and demonstrate its robustness and flexibility in implementation. These results indicate a simple, precise, robust, and scalable quantum control protocol which may greatly facilitate the building of large controlled quantum systems.

\begin{figure}[ht]
	\centering
    \includegraphics[width=\columnwidth]{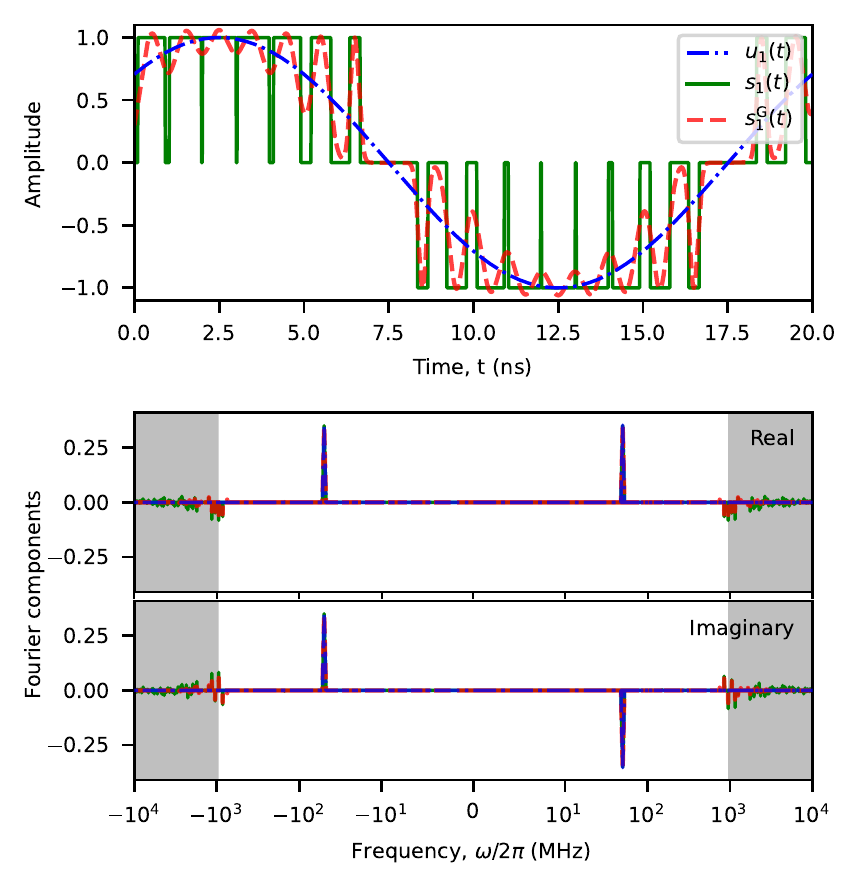}
    \caption{Comparison between a sinusoidal waveform, $u_1(t)=\sin\left(\omega t+\pi/4\right)$ for $\omega/2\pi=50\,{\rm MHz}$ (blue, dash-dotted), and a sequence of rectangular or Gaussian pulses, $s_1(t)$ (green, solid) and $s_1^{\rm G}(t)$ (red, dashed), respectively, in both the time domain and frequency domain. Here, we have used $M=20$ pulses to approximate the waveform in each period. The three functions show almost the same Fourier components below a prescribed threshold $\Omega_1/2\pi \approx 1\,\mathrm{GHz}$. The difference manifests mainly in the high-frequency regime (shaded area).}
    \label{fig:schematic}
\end{figure}

\emph{Waveform-pulse train correspondence.}---We consider a general system with the drift (time-independent) Hamiltonian, $H_0$, and the control (time-dependent) Hamiltonian, $H_c(t)=\sum_{k=1}^{K}u_k(t)H_k$, where $H_k$ is Hermitian and $u_k(t)$ is a real function of time serving as the $k$th control field. Assuming that there exists an optimal waveform, $u_k(t)$, which maximizes a chosen objective, $J(T)$, at the final time, $T$, the key foundation of the PWM method is that there always exists a pulse train, $s_k(t)$, that achieves the same goal to an arbitrary precision \cite{supplementary}. To be specific, we formally define a PWM sequence as
\begin{align}
	s_k\left(t\right)=\sum_{m=1}^{M} \xi_k \left[ \theta\left(t-\tau_{k,m}^{\rm on}\right) - \theta\left(t-\tau_{k,m}^{\rm off}\right) \right],
\end{align}
where $\theta(t)$ is the Heaviside step function, $M=T/\tau$ is the number of pulses in the time interval $[0,T]$. Physically, the pulse train may be generated by operating an ideal switch at time instances $\tau_{k,m}^{\rm on/off}=m\tau-\left(\tau \pm \left|\tau_{k,m}\right|\right)/2$, where 
\begin{align}
	\tau_{k,m}=\int_{(m-1)\tau}^{m\tau} u_k(t) dt/\xi_k. \label{eq:width}
\end{align}
Here, $|\tau_{k,m}|$ represents the pulse width in the $m$th interval, and ${\rm sgn}[\tau_{k,m}]\xi_k$ with $\xi_k = \max|u_k(t)|$ is the amplitude of the pulse train, as illustrated in Fig.~\ref{fig:schematic}. Although the two fields, $u_k(t)$ and $s_k(t)$, are vastly different in the time domain, their Fourier components are almost identical below a prescribed threshold, $\Omega_k\approx 2\pi/\tau$, in frequency. The higher-frequency noise above $\Omega_k$ are relatively small compared with the major frequency components within $\pm\Omega_k$. They can be omitted if the characteristic frequency of the system is much smaller than $\Omega_k$, or may be physically filtered out in experiments. 

To further evaluate the control performance of the pulse train, we use the Dyson series to calculate the average short-time propagation operator for each time interval, $U[m\tau, (m-1)\tau]$, and compare it with the ideal propagator generated by $u_k(t)$. One can prove that the PWM sequence results in a $2$nd-order approximation to the ideal unitary propagation, i.e., $\delta U[m\tau, (m-1)\tau] \propto \mathcal{O}\left( \tau^3 \right)$, which is independent of the number of control degrees of freedom, $K$ \cite{supplementary}. This accuracy is at the same order of the staircase approximation of a waveform generated by AWGs, which reveals the equivalence between a continuous and a pulsed control protocol. 

\emph{Pulsed time propagation.}---To design a PWM sequence for high-fidelity quantum control, one may simply treat the pulse widths, $\tau_{k,m}$, as control variables and resort to numerical QOC algorithms. Besides the continued efforts at minimizing the number of iterations of QOC algorithms, one bottleneck of the optimization is the efficiency of solving the time-dependent Schr{\"o}dinger equation (TDSE). Generally, the latter relies on the calculation of the propagation operator, $U(T,0)$, which involves a considerable number of numerically expensive matrix exponentials during optimization \cite{Moler2003}. However, the PWM approach circumvents this technical challenge by exact matrix decomposition, and may significantly accelerate the calculation. Because $s_k(t)$ can be either \textit{zero} or $\pm \xi_k$, the system Hamiltonian at any time instance is chosen from a finite set 
\begin{align}
	H(t) \in &\left\{H_0, H_0 + \xi_1 H_1, H_0 - \xi_1 H_1, H_0 + \xi_2 H_2,\cdots,\right.\nonumber \\
	&\left.\,
	H_0 + \cdots + \xi_K H_K, \cdots, H_0 - \cdots - \xi_K H_K \right\}.
\end{align}
Thus, one may diagonalize these Hamiltonians in advance and convert the matrix exponentials into scalar exponentials and matrix products. As a concrete example, we consider the simplest case with one unique control field, $K=1$. The propagator for the $m$th time interval can be written as
\begin{widetext}
\begin{align}
	U[m\tau, (m-1)\tau] = \begin{cases}
	P_0\exp\left[-i\left(\tau+\frac{\tau_{1,m}}{2}\right)\Lambda_0\right] P_0^{\dagger}
	P_{-} \exp\left[+i\tau_{1,m}\Lambda_{-}\right] P_{-}^{\dagger}
	P_0 \exp\left[-i\left(\tau+\frac{\tau_{1,m}}{2}\right)\Lambda_0\right] P_0^{\dagger},\,\text{for}\,\tau_{1,m}<0;\\
	P_0\exp\left[-i\left(\tau-\frac{\tau_{1,m}}{2}\right)\Lambda_0\right] P_0^{\dagger}
	P_{+} \exp\left[-i\tau_{1,m}\Lambda_{+}\right] P_{+}^{\dagger}
	P_0 \exp\left[-i\left(\tau-\frac{\tau_{1,m}}{2}\right)\Lambda_0\right] P_0^{\dagger},\,\text{for}\,\tau_{1,m}\geq 0,
	\end{cases}
\end{align}
\end{widetext}
with $H_0 = P_{0}\Lambda_{0}P_{0}^{\dagger}$, $H_0 \pm \xi_1 H_1 = P_{\pm}\Lambda_{\pm}P_{\pm}^{\dagger}$. Here, $\Lambda_{0}$ ($\Lambda_{\pm}$) and $P_{0}$ ($P_{\pm}$) are real diagonal and unitary matrices, respectively, and the $\pm$ sign indicates the polarization of the pulse in the $m$th time interval. This diagonalization process describes the change of basis for representing the matrices, $H_{0}$ and $H_{0}\pm \xi_{1}H_{1}$, such that they are always written in a diagonal form. In each of the bases, the time-propagation induces only a phase factor in the corresponding eigenstates, while the change between the bases captures the possible state transition processes and can be efficiently calculated. This trick is similar to the split operator method for solving TDSE \cite{Fleck1976, Feit1982, Feit1983}. Assuming that the Hilbert space has $d$ dimensions, the above calculation requires approximately $2Kd^3$ floating-point operations (FLOPs) once the decomposition is obtained. By comparison, the typical method of matrix exponentiation, i.e., the scaling and squaring method based on Pad{\'e} approximation, generally requires $10d^3$ to $20d^3$ FLOPs depending on the specific problem and the detailed implementation \cite{Moler2003}. Thus, the pulsed time propagation can be efficiently calculated for a relatively small number of control degrees of freedom, i.e., $K \lesssim 5$ -- $10$. We emphasize that the PWM approach requires $\mathcal{O}\left(d^2\right)$ FLOPs for state propagation problems where the initial state is fixed, which is one-order of magnitude faster than the typical Pad{\'e} approximation method \cite{Moler2003}. 

\emph{Imperfect switches.}---We now study the robustness of the PWM protocol with respect to two imperfect switching events: (i) time jitter and (ii) finite switching speed. For (i), we assume that the $m$th pulse width deviates from the design value by $\delta \tau_{k,m}$, which obeys a Gaussian distribution with mean value \textit{zero} and standard deviation $\sigma_{k}$ \cite{Ding2019}. Then, $\delta \tau_{k,m}\delta \tau_{k',m}/\sigma_{k}\sigma_{k'}$ obeys the $\chi^2$-distribution with expectation value $1$ for $k= k'$, or the generalized Laplace distribution with expectation value \textit{zero} elsewhere. Inserting these relations in the Dyson series of the short-time propagator and keeping only the leading-order terms of $\sigma_{k}$, we estimate the expectation value of the deviation as $\left\langle \delta U[m\tau,(m-1)\tau] \right\rangle \approx \sum_{k=1}^{K}\sigma_{k}^2 \xi_{k}^2H_{k}^2/2$, where $\langle \cdot \rangle$ is the ensemble average. For a qualitative understanding of the error scaling, we assume that $\sigma_{k} = \tau_{k,m}/100$. The time jitter leads to a $0.01\%$ relative error that may be neglected for $T \simeq 10^3\tau$ \footnote{This value is estimated by noticing that $0.9999^{1000} \approx 0.9$.}.

To account for the influence of a finite switching speed, (ii), we consider a sequence of Gaussian pulses 
\begin{align}
	s_k^{\rm G}(t) = \sum_{m=1}^{M}\xi_k e^{-\pi\left[t-(m-\frac{1}{2})\tau\right]^2/\tau_{k,m}^2},
\end{align}
where $\tau_{k,m}$ is defined in Eq.\,\eqref{eq:width}. The bell-like edges and the overlap between adjacent pulses describe a relatively slow switch, which distorts the desired rectangular pulses. However, one can prove that the Fourier components of $s_k^{\rm G}(t)$ is identical to the ideal case below $\Omega_{k}$ \cite{supplementary}, as illustrated in Fig.\,\ref{fig:schematic}. Thus, the control performance has the same order of precision as the perfect rectangular pulses.

\emph{Generalizations.}---Following the same argument of waveform-pulse train correspondence, one can generalize the PWM control protocol in a variety ways, for example, by removing the $s_k(t)=0$ stage of the sequence. The resulting binary sequence is called the $2$-level PWM pulse sequence in the literature, which keeps the signal power unchanged and alters only the polarization of the switch. Similarly, one can also define a $n$-level PWM sequence with $n$ discrete polarizations of $s_k(t)$. From another perspective, one may stretch the pulse magnitude, $\xi_k$, by a factor of $\kappa$ and shrink the pulse width by $1/\kappa$ at the same time, while keeping the control performance almost unchanged. The extreme case with $\kappa \rightarrow \infty$ represents a sequence of hard pulses for quantum control, where the control protocol applies a sequence of instantaneous but non-negligible kicks to the system \cite{Viola1999, Viola1999a, Lloyd2001}. Here, the corresponding time propagation is identical to the symmetrically decomposed Suzuki-Trotter formula \cite{Trotter1959, Suzuki1985}
\begin{equation}
	U[m\tau,(m-1)\tau] = \left(\prod_{k=0}^{K} e^{\frac{-i\tau_k}{2}H_k}\right) 
	\left(\prod_{k=K}^{0} e^{\frac{-i\tau_k}{2}H_k}\right), 
	\label{eq:spo}
\end{equation}
where $\tau_{0,m}=\tau$ and $\tau_{k,m}$ is the pulse width defined in Eq.\,\eqref{eq:width}. Detailed analysis shows that Eq.\,\eqref{eq:spo} has a slightly larger error rate than the $\kappa=1$ case, as the high-frequency components above $\Omega_{k}$ have larger magnitudes \cite{supplementary}. However, the system Hamiltonian at any time instance now belongs to a smaller set, $H(t)\in\{H_0, \pm\xi_1 H_1, \cdots \pm\xi_{K} H_{K}\}$, which makes the simulation of the TDSE even more efficient without substantially deteriorating the numerical precision.

Using the high-order form of the Suzuki-Trotter formula \cite{Bandrauk1992, Bandrauk1993}, one can insert more pulses in each time interval and increase the precision for solving TDSE and also for physical implementations. One can prove that the $(2n+1)$th-order-accurate PWM sequence, with all the pulse widths being real and positive, always exists \cite{supplementary}. The corresponding number of pulses in each interval is $3^{2n-1}$, where the new pulse widths can be derived from the $2$nd-order-accurate form \cite{supplementary}. By comparison, the accuracy of the staircase approximation remains at the $2$nd order when splitting each short time interval by the same number pieces.  

\begin{figure}
  \centering
  \includegraphics[width=\columnwidth]{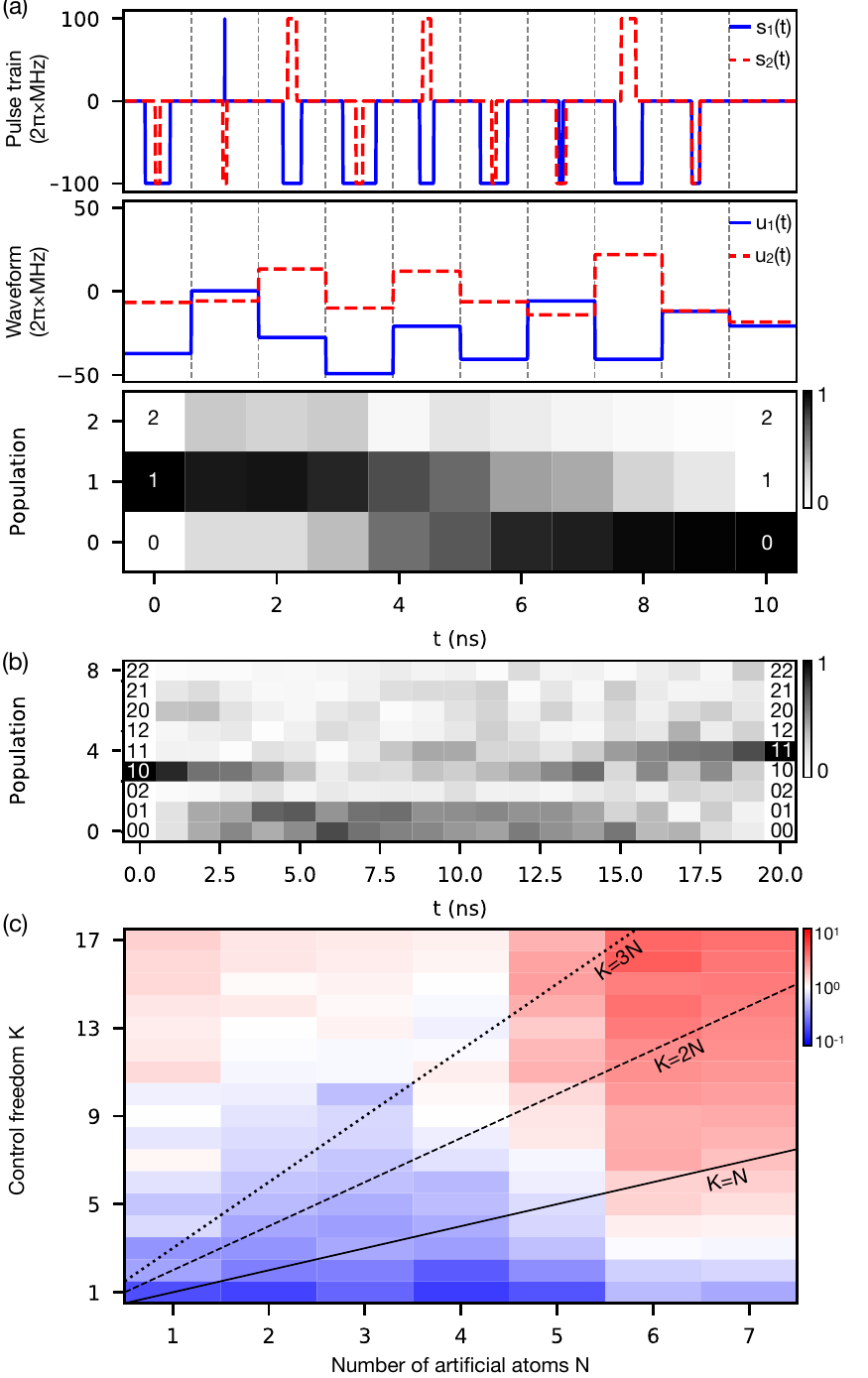}
  \caption{(a) The optimized single-qubit control pulse train (top) and the corresponding staircase waveform (middle). The bottom panel shows the dynamics of the system during the control process. (b) The population of different states during an optimized CNOT gate. In both (a) and (b), the ternary string indicates the energy level of the corresponding artificial atoms. (c) Comparison of the computational resources, $\gamma$, between the PWM and the conventional approaches for solving TDSE. The former outperforms the latter for a relatively small control freedom, $K \lesssim 5$ -- $10$, which is consistent with the FLOP analysis.} 
  \label{fig:simulation}
\end{figure}

\emph{Application to SQC.}---For illustration, we consider a chain of superconducting artificial atoms with XY-control, $u_{n,x/y}(t)$ for $n=1,\cdots,N$. Here, each artificial atom is modeled as a Kerr-nonlinear resonator with annihilation and creation operators, $a_{n}$ and $a_{n}^{\dagger}$, of which the lowest two energy levels, $|0_{n}\rangle$ and $|1_{n}\rangle$, are encoded as a qubit. In the doubly rotating frame at the qubit frequency, the total Hamiltonian reads \cite{Galiautdinov2012, Ghosh2013, Barends2014, Kelly2015, Roushan2017}
\begin{align}
	H &= \sum_{n=1}^{N} \frac{\eta_{n}}{2}a_{n}^{\dagger}a_{n}^{\dagger}a_{n}a_{n}
	+ \sum_{n=1}^{N-1} g_{n,n+1} \left( a_{n}^{\dagger}a_{n+1} + a_{n}a_{n+1}^{\dagger} \right)\nonumber \\
	&+ \sum_{n=1}^{N} \left[ u_{n,x}(t)\left(a_n+a_n^{\dagger}\right)
	+ iu_{n,y}(t)\left(a_n-a_n^{\dagger}\right) \right]
\end{align}
For simplicity, we truncate the Hilbert space of each artificial atom to $3$ dimensions, and assume that all the atoms are homogeneous with anharmonicity $\eta_{n}/2\pi=-200\,{\rm MHz}$ and coupling strength $g_{n,n+1}/2\pi=30\,{\rm MHz}$. We evaluate the control performance by the average fidelity over all the relevant qubit states \cite{Zanardi2004, Pedersen2007}
\begin{align}
	J(T) = \frac{{\rm tr}\left[U(T,0)PU^{\dagger}(T,0)P\right] + \left|{\rm tr}\left[U_{\rm g}^{\dagger}U(T,0)P\right]\right|^2}{2^{N}\left(2^{N}+1\right)},
\end{align}
where $P=\otimes_{n=1}^{N}\left(\mathbbm{1}_{n}-|2_n\rangle\langle 2_n|\right)$ is the projection operator onto the qubit subspace, and $U_{\rm g}$ is the target gate. 

We study first the $N=1$ case with $T=10\,{\rm ns}$ and $\xi_{1,x/y}/2\pi = 100\,{\rm MHz}$. Our goal is to implement a NOT gate while preventing population leakage from the qubit subspace to higher energy levels \cite{Tian2000, Steffen2003, Motzoi2009, Safaei2009, Rebentrost2009, Schutjens2013}. The optimized pulse trains and the converted staircase waveforms are shown in Fig.\,\ref{fig:simulation}(a) top and middle, respectively, of which the fidelities are $0.9999$ and $0.9997$ \footnote{We note that all the converted waveforms for AWG implementation can be further optimized to $0.9999$ fidelity with conventional QOC algorithms.}. Here, the conversion is achieved by inverting Eq.\,\eqref{eq:width}, i.e., $u_k(m\tau) = \xi_k \tau_{k,m}/\tau$, with $\tau=1\,{\rm ns}$ time-resolution for the waveform. This result clearly indicates the correspondence between an arbitrary waveform and a pulse train, which is the basis of the PWM method. Furthermore, it demonstrates that a sequence of properly concatenated pulses can eliminate leakage in controlling a superconducting qubit, although each single pulse is broad in frequency and covers multiple energy levels. The detailed transfer among the three energy levels during the control process is illustrated in Fig.\,\ref{fig:simulation}(a) bottom, where the system is initially in $|1_{1}\rangle$. 

Next, we consider the implementation of a CNOT gate with $N=2$ and $\xi_{n,x/y}/2\pi = 100\,{\rm MHz}$. The fidelity is optimized to $0.9999$ using pulse trains, while that of the converted waveforms is $0.9848$. Similarly, we illustrate the dynamics of the system in Fig.\,\ref{fig:simulation}(b), where the initial state $|1_{1}0_{2}\rangle$ is transferred to $|1_{1}1_{2}\rangle$ at $T=20\,{\rm ns}$. We also optimized the pulse trains for implementing a $0.9999$-fidelity CCZ gate at $T=30\,{\rm ns}$ with three $\pm 700\,{\rm MHz}$-range Z-controls. The fidelity of the converted waveform is $0.9195$. These examples demonstrate the potential of using a PWM control protocol for realizing the universal control of superconducting qubits. 

For more complex systems, we compare the computational efficiency of the PWM and the conventional approaches for solving TDSE \cite{supplementary}, as shown in Fig.\,\ref{fig:simulation}(c). Here, we vary the control freedom, $K$, and system size, $N$, and repeat the simulation for $10$ times with random control fields, $u_k(t)$ or $s_k(t)$, for average performance. The PWM method outperforms the default method for $N\leq 4$, $K\lesssim 10$ and $N \geq 5$, $K\lesssim 5$, which is qualitatively consistent with the FLOP analysis. Over the simulated parameter regime, the two corresponding numerical time consumptions achieve a minimum ratio of $\gamma=0.17$ at $N=4$, $K=1$, and a maximum of $4.36$ at $N=6$, $K=16$. The average value below $K=5$ (included) is $0.57$, which indicates a significant acceleration of the numerical calculation. 

For an initial proof-of-principle demonstration of the PWM controller, we expect the use of high-speed room-temperature AWGs to generate the required pulse sequence. Here, a high sampling rate is required because of the unevenly spaced switching time instances, which is the drawback of the PWM approach compared with AWG. Towards a scalable low-temperature integration of the controller, we anticipate that the PWM controller can be readily realized with current cryogenic complementary metal-oxide-semiconductor (cryo-CMOS) \cite{Hornibrook2015, Reilly2015, Vandersypen2017, Bardin2019, Dijk2019, Petit2020, Pauka2021, Xue2021} or rapid single flux quantum (RSFQ) technologies \cite{Likharev1991, Zhou2001, Crankshaw2003, Semenov2003}. They have a typical switching speed of $\sim 100\,{\rm ps}$ and $\sim 1\,{\rm ps}$, respectively. Both technologies enable the relocation of the controller to the cryogenic environment, which avoids racks of room-temperature electronics and substantially reduces the need for lossy and noisy cables connecting the controller and the quantum processor. One may either mix a relatively slow pulse sequence generated by the cryo-CMOS switch with a fast oscillating carrier wave, or generate a sequence of rapid PWM pulses directly with RSFQ switches. Compared with a cryogenic AWG, where tens of cryo-COMS switches are superimposed to generate an arbitrary waveform, the PWM controller requires only a single switch for each degree of control freedom and thus leads to dramatic reduction of the hardware demands. Compared with the existing RSFQ approach \cite{McDermott2014, Liebermann2016, McDermott2018, Leonard2019, Li2019}, the PWM pulses naturally have a finite width. It thus indicates a slower switching rate rather than generating a sequence of $\sim 2\,{\rm ps}$-width RSFQ pulses. The number of switching events is also substantially smaller than the RSFQ approach. 

\emph{Conclusions and outlook.}---We apply the concept of PWM to quantum systems, and propose the PWM method that enables arbitrary control of a general quantum system with a sequence of simple pulses. The performance of the pulse train can be efficiently evaluated by a matrix decomposition method, which avoids numerically expensive matrix exponentials and accelerates the TDSE solver for a relatively small number of controls. The control protocol is robust to time jitter, switching delay, and leakage. The implementations in SQC may be readily achieved with existing semiconductor or superconductor technologies but with a lower demand of hardware resources. These results indicate a simple, precise, robust, and scalable control protocol for a general quantum system. Together with the advances of high-density wiring and packaging techniques, an experimental realization of the PWM controller at cryogenic temperature may offer significant advances for building a practically useful quantum information processor.

Besides achieving high fidelity, one further goal in QOC is to implement the controls as fast as possible. Recent experiments have demonstrated that, in spin systems, the time-optimal control fields may converge to a pulse sequence form when $T$ approaches its theoretical minimum value \cite{Khaneja2001, Chen2015, Chen2020}. This observation indicates that a pulse train might be a more suitable and natural choice than a continuous waveform for solving quantum time-optimal control problems. We anticipate that the PWM method may open up new ways for seeking time-optimal control pulses in a general quantum system.

\emph{Acknowledgements.}---We thank Benjamin Lienhard, Olli-Pentti Saira, and Visa Vesterinen for insightful discussions. R.W. acknowledges support from National Natural Science Foundation of China (Grant No.\,62173201 and No.\,61833010). H.R. acknowledges support from the U.S. Army Research Office (Grant No.\,W911NF-19-1-0382). The codes that support the findings of this study are available at https://github.com/chenqmion/QPWM.

\bibliography{PWM_REF}  

\begin{thebibliography}{66}%
\makeatletter
\providecommand \@ifxundefined [1]{%
 \@ifx{#1\undefined}
}%
\providecommand \@ifnum [1]{%
 \ifnum #1\expandafter \@firstoftwo
 \else \expandafter \@secondoftwo
 \fi
}%
\providecommand \@ifx [1]{%
 \ifx #1\expandafter \@firstoftwo
 \else \expandafter \@secondoftwo
 \fi
}%
\providecommand \natexlab [1]{#1}%
\providecommand \enquote  [1]{``#1''}%
\providecommand \bibnamefont  [1]{#1}%
\providecommand \bibfnamefont [1]{#1}%
\providecommand \citenamefont [1]{#1}%
\providecommand \href@noop [0]{\@secondoftwo}%
\providecommand \href [0]{\begingroup \@sanitize@url \@href}%
\providecommand \@href[1]{\@@startlink{#1}\@@href}%
\providecommand \@@href[1]{\endgroup#1\@@endlink}%
\providecommand \@sanitize@url [0]{\catcode `\\12\catcode `\$12\catcode
  `\&12\catcode `\#12\catcode `\^12\catcode `\_12\catcode `\%12\relax}%
\providecommand \@@startlink[1]{}%
\providecommand \@@endlink[0]{}%
\providecommand \url  [0]{\begingroup\@sanitize@url \@url }%
\providecommand \@url [1]{\endgroup\@href {#1}{\urlprefix }}%
\providecommand \urlprefix  [0]{URL }%
\providecommand \Eprint [0]{\href }%
\providecommand \doibase [0]{https://doi.org/}%
\providecommand \selectlanguage [0]{\@gobble}%
\providecommand \bibinfo  [0]{\@secondoftwo}%
\providecommand \bibfield  [0]{\@secondoftwo}%
\providecommand \translation [1]{[#1]}%
\providecommand \BibitemOpen [0]{}%
\providecommand \bibitemStop [0]{}%
\providecommand \bibitemNoStop [0]{.\EOS\space}%
\providecommand \EOS [0]{\spacefactor3000\relax}%
\providecommand \BibitemShut  [1]{\csname bibitem#1\endcsname}%
\let\auto@bib@innerbib\@empty
\bibitem [{\citenamefont {Warren}\ \emph {et~al.}(1993)\citenamefont {Warren},
  \citenamefont {Rabitz},\ and\ \citenamefont {Dahleh}}]{Warren1993}%
  \BibitemOpen
  \bibfield  {author} {\bibinfo {author} {\bibfnamefont {W.~S.}\ \bibnamefont
  {Warren}}, \bibinfo {author} {\bibfnamefont {H.}~\bibnamefont {Rabitz}},\
  and\ \bibinfo {author} {\bibfnamefont {M.}~\bibnamefont {Dahleh}},\
  }\bibfield  {title} {\bibinfo {title} {Coherent control of quantum dynamics:
  The dream is alive},\ }\href {https://doi.org/10.1126/science.259.5101.1581}
  {\bibfield  {journal} {\bibinfo  {journal} {Science}\ }\textbf {\bibinfo
  {volume} {259}},\ \bibinfo {pages} {1581} (\bibinfo {year}
  {1993})}\BibitemShut {NoStop}%
\bibitem [{\citenamefont {Rabitz}\ \emph {et~al.}(2000)\citenamefont {Rabitz},
  \citenamefont {de~Vivie-Riedle}, \citenamefont {Motzkus},\ and\ \citenamefont
  {Kompa}}]{Rabitz2000}%
  \BibitemOpen
  \bibfield  {author} {\bibinfo {author} {\bibfnamefont {H.}~\bibnamefont
  {Rabitz}}, \bibinfo {author} {\bibfnamefont {R.}~\bibnamefont
  {de~Vivie-Riedle}}, \bibinfo {author} {\bibfnamefont {M.}~\bibnamefont
  {Motzkus}},\ and\ \bibinfo {author} {\bibfnamefont {K.}~\bibnamefont
  {Kompa}},\ }\bibfield  {title} {\bibinfo {title} {Whither the future of
  controlling quantum phenomena?},\ }\href
  {https://doi.org/10.1126/science.288.5467.824} {\bibfield  {journal}
  {\bibinfo  {journal} {Science}\ }\textbf {\bibinfo {volume} {288}},\ \bibinfo
  {pages} {824} (\bibinfo {year} {2000})}\BibitemShut {NoStop}%
\bibitem [{\citenamefont {Peirce}\ \emph {et~al.}(1988)\citenamefont {Peirce},
  \citenamefont {Dahleh},\ and\ \citenamefont {Rabitz}}]{Peirce1988}%
  \BibitemOpen
  \bibfield  {author} {\bibinfo {author} {\bibfnamefont {A.~P.}\ \bibnamefont
  {Peirce}}, \bibinfo {author} {\bibfnamefont {M.~A.}\ \bibnamefont {Dahleh}},\
  and\ \bibinfo {author} {\bibfnamefont {H.}~\bibnamefont {Rabitz}},\
  }\bibfield  {title} {\bibinfo {title} {Optimal control of quantum-mechanical
  systems: Existence, numerical approximation, and applications},\ }\href
  {https://doi.org/10.1103/PhysRevA.37.4950} {\bibfield  {journal} {\bibinfo
  {journal} {Phys. Rev. A}\ }\textbf {\bibinfo {volume} {37}},\ \bibinfo
  {pages} {4950} (\bibinfo {year} {1988})}\BibitemShut {NoStop}%
\bibitem [{\citenamefont {Kosloff}\ \emph {et~al.}(1989)\citenamefont
  {Kosloff}, \citenamefont {Rice}, \citenamefont {Gaspard}, \citenamefont
  {Tersigni},\ and\ \citenamefont {Tannor}}]{Kosloff1989}%
  \BibitemOpen
  \bibfield  {author} {\bibinfo {author} {\bibfnamefont {R.}~\bibnamefont
  {Kosloff}}, \bibinfo {author} {\bibfnamefont {S.}~\bibnamefont {Rice}},
  \bibinfo {author} {\bibfnamefont {P.}~\bibnamefont {Gaspard}}, \bibinfo
  {author} {\bibfnamefont {S.}~\bibnamefont {Tersigni}},\ and\ \bibinfo
  {author} {\bibfnamefont {D.}~\bibnamefont {Tannor}},\ }\bibfield  {title}
  {\bibinfo {title} {Wavepacket dancing: Achieving chemical selectivity by
  shaping light pulses},\ }\href
  {https://doi.org/https://doi.org/10.1016/0301-0104(89)90012-8} {\bibfield
  {journal} {\bibinfo  {journal} {Chem. Phys.}\ }\textbf {\bibinfo {volume}
  {139}},\ \bibinfo {pages} {201 } (\bibinfo {year} {1989})}\BibitemShut
  {NoStop}%
\bibitem [{\citenamefont {Judson}\ and\ \citenamefont
  {Rabitz}(1992)}]{Judson1992}%
  \BibitemOpen
  \bibfield  {author} {\bibinfo {author} {\bibfnamefont {R.~S.}\ \bibnamefont
  {Judson}}\ and\ \bibinfo {author} {\bibfnamefont {H.}~\bibnamefont
  {Rabitz}},\ }\bibfield  {title} {\bibinfo {title} {Teaching lasers to control
  molecules},\ }\href {https://doi.org/10.1103/PhysRevLett.68.1500} {\bibfield
  {journal} {\bibinfo  {journal} {Phys. Rev. Lett.}\ }\textbf {\bibinfo
  {volume} {68}},\ \bibinfo {pages} {1500} (\bibinfo {year}
  {1992})}\BibitemShut {NoStop}%
\bibitem [{\citenamefont {Assion}\ \emph {et~al.}(1998)\citenamefont {Assion},
  \citenamefont {Baumert}, \citenamefont {Bergt}, \citenamefont {Brixner},
  \citenamefont {Kiefer}, \citenamefont {Seyfried}, \citenamefont {Strehle},\
  and\ \citenamefont {Gerber}}]{Assion1998}%
  \BibitemOpen
  \bibfield  {author} {\bibinfo {author} {\bibfnamefont {A.}~\bibnamefont
  {Assion}}, \bibinfo {author} {\bibfnamefont {T.}~\bibnamefont {Baumert}},
  \bibinfo {author} {\bibfnamefont {M.}~\bibnamefont {Bergt}}, \bibinfo
  {author} {\bibfnamefont {T.}~\bibnamefont {Brixner}}, \bibinfo {author}
  {\bibfnamefont {B.}~\bibnamefont {Kiefer}}, \bibinfo {author} {\bibfnamefont
  {V.}~\bibnamefont {Seyfried}}, \bibinfo {author} {\bibfnamefont
  {M.}~\bibnamefont {Strehle}},\ and\ \bibinfo {author} {\bibfnamefont
  {G.}~\bibnamefont {Gerber}},\ }\bibfield  {title} {\bibinfo {title} {Control
  of chemical reactions by feedback-optimized phase-shaped femtosecond laser
  pulses},\ }\href {https://doi.org/10.1126/science.282.5390.919} {\bibfield
  {journal} {\bibinfo  {journal} {Science}\ }\textbf {\bibinfo {volume}
  {282}},\ \bibinfo {pages} {919} (\bibinfo {year} {1998})}\BibitemShut
  {NoStop}%
\bibitem [{\citenamefont {Tian}\ and\ \citenamefont {Lloyd}(2000)}]{Tian2000}%
  \BibitemOpen
  \bibfield  {author} {\bibinfo {author} {\bibfnamefont {L.}~\bibnamefont
  {Tian}}\ and\ \bibinfo {author} {\bibfnamefont {S.}~\bibnamefont {Lloyd}},\
  }\bibfield  {title} {\bibinfo {title} {Resonant cancellation of off-resonant
  effects in a multilevel qubit},\ }\href
  {https://doi.org/10.1103/PhysRevA.62.050301} {\bibfield  {journal} {\bibinfo
  {journal} {Phys. Rev. A}\ }\textbf {\bibinfo {volume} {62}},\ \bibinfo
  {pages} {050301} (\bibinfo {year} {2000})}\BibitemShut {NoStop}%
\bibitem [{\citenamefont {Steffen}\ \emph {et~al.}(2003)\citenamefont
  {Steffen}, \citenamefont {Martinis},\ and\ \citenamefont
  {Chuang}}]{Steffen2003}%
  \BibitemOpen
  \bibfield  {author} {\bibinfo {author} {\bibfnamefont {M.}~\bibnamefont
  {Steffen}}, \bibinfo {author} {\bibfnamefont {J.~M.}\ \bibnamefont
  {Martinis}},\ and\ \bibinfo {author} {\bibfnamefont {I.~L.}\ \bibnamefont
  {Chuang}},\ }\bibfield  {title} {\bibinfo {title} {Accurate control of
  josephson phase qubits},\ }\href {https://doi.org/10.1103/PhysRevB.68.224518}
  {\bibfield  {journal} {\bibinfo  {journal} {Phys. Rev. B}\ }\textbf {\bibinfo
  {volume} {68}},\ \bibinfo {pages} {224518} (\bibinfo {year}
  {2003})}\BibitemShut {NoStop}%
\bibitem [{\citenamefont {Motzoi}\ \emph {et~al.}(2009)\citenamefont {Motzoi},
  \citenamefont {Gambetta}, \citenamefont {Rebentrost},\ and\ \citenamefont
  {Wilhelm}}]{Motzoi2009}%
  \BibitemOpen
  \bibfield  {author} {\bibinfo {author} {\bibfnamefont {F.}~\bibnamefont
  {Motzoi}}, \bibinfo {author} {\bibfnamefont {J.~M.}\ \bibnamefont
  {Gambetta}}, \bibinfo {author} {\bibfnamefont {P.}~\bibnamefont
  {Rebentrost}},\ and\ \bibinfo {author} {\bibfnamefont {F.~K.}\ \bibnamefont
  {Wilhelm}},\ }\bibfield  {title} {\bibinfo {title} {Simple pulses for
  elimination of leakage in weakly nonlinear qubits},\ }\href
  {https://doi.org/10.1103/PhysRevLett.103.110501} {\bibfield  {journal}
  {\bibinfo  {journal} {Phys. Rev. Lett.}\ }\textbf {\bibinfo {volume} {103}},\
  \bibinfo {pages} {110501} (\bibinfo {year} {2009})}\BibitemShut {NoStop}%
\bibitem [{\citenamefont {Safaei}\ \emph {et~al.}(2009)\citenamefont {Safaei},
  \citenamefont {Montangero}, \citenamefont {Taddei},\ and\ \citenamefont
  {Fazio}}]{Safaei2009}%
  \BibitemOpen
  \bibfield  {author} {\bibinfo {author} {\bibfnamefont {S.}~\bibnamefont
  {Safaei}}, \bibinfo {author} {\bibfnamefont {S.}~\bibnamefont {Montangero}},
  \bibinfo {author} {\bibfnamefont {F.}~\bibnamefont {Taddei}},\ and\ \bibinfo
  {author} {\bibfnamefont {R.}~\bibnamefont {Fazio}},\ }\bibfield  {title}
  {\bibinfo {title} {Optimized single-qubit gates for josephson phase qubits},\
  }\href {https://doi.org/10.1103/PhysRevB.79.064524} {\bibfield  {journal}
  {\bibinfo  {journal} {Phys. Rev. B}\ }\textbf {\bibinfo {volume} {79}},\
  \bibinfo {pages} {064524} (\bibinfo {year} {2009})}\BibitemShut {NoStop}%
\bibitem [{\citenamefont {Rebentrost}\ and\ \citenamefont
  {Wilhelm}(2009)}]{Rebentrost2009}%
  \BibitemOpen
  \bibfield  {author} {\bibinfo {author} {\bibfnamefont {P.}~\bibnamefont
  {Rebentrost}}\ and\ \bibinfo {author} {\bibfnamefont {F.~K.}\ \bibnamefont
  {Wilhelm}},\ }\bibfield  {title} {\bibinfo {title} {Optimal control of a
  leaking qubit},\ }\href {https://doi.org/10.1103/PhysRevB.79.060507}
  {\bibfield  {journal} {\bibinfo  {journal} {Phys. Rev. B}\ }\textbf {\bibinfo
  {volume} {79}},\ \bibinfo {pages} {060507} (\bibinfo {year}
  {2009})}\BibitemShut {NoStop}%
\bibitem [{\citenamefont {Schutjens}\ \emph {et~al.}(2013)\citenamefont
  {Schutjens}, \citenamefont {Dagga}, \citenamefont {Egger},\ and\
  \citenamefont {Wilhelm}}]{Schutjens2013}%
  \BibitemOpen
  \bibfield  {author} {\bibinfo {author} {\bibfnamefont {R.}~\bibnamefont
  {Schutjens}}, \bibinfo {author} {\bibfnamefont {F.~A.}\ \bibnamefont
  {Dagga}}, \bibinfo {author} {\bibfnamefont {D.~J.}\ \bibnamefont {Egger}},\
  and\ \bibinfo {author} {\bibfnamefont {F.~K.}\ \bibnamefont {Wilhelm}},\
  }\bibfield  {title} {\bibinfo {title} {Single-qubit gates in
  frequency-crowded transmon systems},\ }\href
  {https://doi.org/10.1103/PhysRevA.88.052330} {\bibfield  {journal} {\bibinfo
  {journal} {Phys. Rev. A}\ }\textbf {\bibinfo {volume} {88}},\ \bibinfo
  {pages} {052330} (\bibinfo {year} {2013})}\BibitemShut {NoStop}%
\bibitem [{\citenamefont {Zahedinejad}\ \emph {et~al.}(2015)\citenamefont
  {Zahedinejad}, \citenamefont {Ghosh},\ and\ \citenamefont
  {Sanders}}]{Zahedinejad2015}%
  \BibitemOpen
  \bibfield  {author} {\bibinfo {author} {\bibfnamefont {E.}~\bibnamefont
  {Zahedinejad}}, \bibinfo {author} {\bibfnamefont {J.}~\bibnamefont {Ghosh}},\
  and\ \bibinfo {author} {\bibfnamefont {B.~C.}\ \bibnamefont {Sanders}},\
  }\bibfield  {title} {\bibinfo {title} {High-fidelity single-shot toffoli gate
  via quantum control},\ }\href
  {https://doi.org/10.1103/PhysRevLett.114.200502} {\bibfield  {journal}
  {\bibinfo  {journal} {Phys. Rev. Lett.}\ }\textbf {\bibinfo {volume} {114}},\
  \bibinfo {pages} {200502} (\bibinfo {year} {2015})}\BibitemShut {NoStop}%
\bibitem [{\citenamefont {Chow}\ \emph {et~al.}(2010)\citenamefont {Chow},
  \citenamefont {DiCarlo}, \citenamefont {Gambetta}, \citenamefont {Motzoi},
  \citenamefont {Frunzio}, \citenamefont {Girvin},\ and\ \citenamefont
  {Schoelkopf}}]{Chow2010}%
  \BibitemOpen
  \bibfield  {author} {\bibinfo {author} {\bibfnamefont {J.~M.}\ \bibnamefont
  {Chow}}, \bibinfo {author} {\bibfnamefont {L.}~\bibnamefont {DiCarlo}},
  \bibinfo {author} {\bibfnamefont {J.~M.}\ \bibnamefont {Gambetta}}, \bibinfo
  {author} {\bibfnamefont {F.}~\bibnamefont {Motzoi}}, \bibinfo {author}
  {\bibfnamefont {L.}~\bibnamefont {Frunzio}}, \bibinfo {author} {\bibfnamefont
  {S.~M.}\ \bibnamefont {Girvin}},\ and\ \bibinfo {author} {\bibfnamefont
  {R.~J.}\ \bibnamefont {Schoelkopf}},\ }\bibfield  {title} {\bibinfo {title}
  {Optimized driving of superconducting artificial atoms for improved
  single-qubit gates},\ }\href {https://doi.org/10.1103/PhysRevA.82.040305}
  {\bibfield  {journal} {\bibinfo  {journal} {Phys. Rev. A}\ }\textbf {\bibinfo
  {volume} {82}},\ \bibinfo {pages} {040305} (\bibinfo {year}
  {2010})}\BibitemShut {NoStop}%
\bibitem [{\citenamefont {Chow}\ \emph {et~al.}(2011)\citenamefont {Chow},
  \citenamefont {C\'orcoles}, \citenamefont {Gambetta}, \citenamefont
  {Rigetti}, \citenamefont {Johnson}, \citenamefont {Smolin}, \citenamefont
  {Rozen}, \citenamefont {Keefe}, \citenamefont {Rothwell}, \citenamefont
  {Ketchen},\ and\ \citenamefont {Steffen}}]{Chow2011}%
  \BibitemOpen
  \bibfield  {author} {\bibinfo {author} {\bibfnamefont {J.~M.}\ \bibnamefont
  {Chow}}, \bibinfo {author} {\bibfnamefont {A.~D.}\ \bibnamefont
  {C\'orcoles}}, \bibinfo {author} {\bibfnamefont {J.~M.}\ \bibnamefont
  {Gambetta}}, \bibinfo {author} {\bibfnamefont {C.}~\bibnamefont {Rigetti}},
  \bibinfo {author} {\bibfnamefont {B.~R.}\ \bibnamefont {Johnson}}, \bibinfo
  {author} {\bibfnamefont {J.~A.}\ \bibnamefont {Smolin}}, \bibinfo {author}
  {\bibfnamefont {J.~R.}\ \bibnamefont {Rozen}}, \bibinfo {author}
  {\bibfnamefont {G.~A.}\ \bibnamefont {Keefe}}, \bibinfo {author}
  {\bibfnamefont {M.~B.}\ \bibnamefont {Rothwell}}, \bibinfo {author}
  {\bibfnamefont {M.~B.}\ \bibnamefont {Ketchen}},\ and\ \bibinfo {author}
  {\bibfnamefont {M.}~\bibnamefont {Steffen}},\ }\bibfield  {title} {\bibinfo
  {title} {Simple all-microwave entangling gate for fixed-frequency
  superconducting qubits},\ }\href
  {https://doi.org/10.1103/PhysRevLett.107.080502} {\bibfield  {journal}
  {\bibinfo  {journal} {Phys. Rev. Lett.}\ }\textbf {\bibinfo {volume} {107}},\
  \bibinfo {pages} {080502} (\bibinfo {year} {2011})}\BibitemShut {NoStop}%
\bibitem [{\citenamefont {Chow}\ \emph {et~al.}(2012)\citenamefont {Chow},
  \citenamefont {Gambetta}, \citenamefont {C\'orcoles}, \citenamefont {Merkel},
  \citenamefont {Smolin}, \citenamefont {Rigetti}, \citenamefont {Poletto},
  \citenamefont {Keefe}, \citenamefont {Rothwell}, \citenamefont {Rozen},
  \citenamefont {Ketchen},\ and\ \citenamefont {Steffen}}]{Chow2012}%
  \BibitemOpen
  \bibfield  {author} {\bibinfo {author} {\bibfnamefont {J.~M.}\ \bibnamefont
  {Chow}}, \bibinfo {author} {\bibfnamefont {J.~M.}\ \bibnamefont {Gambetta}},
  \bibinfo {author} {\bibfnamefont {A.~D.}\ \bibnamefont {C\'orcoles}},
  \bibinfo {author} {\bibfnamefont {S.~T.}\ \bibnamefont {Merkel}}, \bibinfo
  {author} {\bibfnamefont {J.~A.}\ \bibnamefont {Smolin}}, \bibinfo {author}
  {\bibfnamefont {C.}~\bibnamefont {Rigetti}}, \bibinfo {author} {\bibfnamefont
  {S.}~\bibnamefont {Poletto}}, \bibinfo {author} {\bibfnamefont {G.~A.}\
  \bibnamefont {Keefe}}, \bibinfo {author} {\bibfnamefont {M.~B.}\ \bibnamefont
  {Rothwell}}, \bibinfo {author} {\bibfnamefont {J.~R.}\ \bibnamefont {Rozen}},
  \bibinfo {author} {\bibfnamefont {M.~B.}\ \bibnamefont {Ketchen}},\ and\
  \bibinfo {author} {\bibfnamefont {M.}~\bibnamefont {Steffen}},\ }\bibfield
  {title} {\bibinfo {title} {Universal quantum gate set approaching
  fault-tolerant thresholds with superconducting qubits},\ }\href
  {https://doi.org/10.1103/PhysRevLett.109.060501} {\bibfield  {journal}
  {\bibinfo  {journal} {Phys. Rev. Lett.}\ }\textbf {\bibinfo {volume} {109}},\
  \bibinfo {pages} {060501} (\bibinfo {year} {2012})}\BibitemShut {NoStop}%
\bibitem [{\citenamefont {Heeres}\ \emph {et~al.}(2017)\citenamefont {Heeres},
  \citenamefont {Reinhold}, \citenamefont {Ofek}, \citenamefont {Frunzio},
  \citenamefont {Jiang}, \citenamefont {Devoret},\ and\ \citenamefont
  {Schoelkopf}}]{Heeres2017}%
  \BibitemOpen
  \bibfield  {author} {\bibinfo {author} {\bibfnamefont {R.~W.}\ \bibnamefont
  {Heeres}}, \bibinfo {author} {\bibfnamefont {P.}~\bibnamefont {Reinhold}},
  \bibinfo {author} {\bibfnamefont {N.}~\bibnamefont {Ofek}}, \bibinfo {author}
  {\bibfnamefont {L.}~\bibnamefont {Frunzio}}, \bibinfo {author} {\bibfnamefont
  {L.}~\bibnamefont {Jiang}}, \bibinfo {author} {\bibfnamefont {M.~H.}\
  \bibnamefont {Devoret}},\ and\ \bibinfo {author} {\bibfnamefont {R.~J.}\
  \bibnamefont {Schoelkopf}},\ }\bibfield  {title} {\bibinfo {title}
  {Implementing a universal gate set on a logical qubit encoded in an
  oscillator},\ }\href
  {https://doi.org/https://doi.org/10.1038/s41467-017-00045-1} {\bibfield
  {journal} {\bibinfo  {journal} {Nat. Commun.}\ }\textbf {\bibinfo {volume}
  {8}},\ \bibinfo {pages} {94} (\bibinfo {year} {2017})}\BibitemShut {NoStop}%
\bibitem [{\citenamefont {Galiautdinov}\ \emph {et~al.}(2012)\citenamefont
  {Galiautdinov}, \citenamefont {Korotkov},\ and\ \citenamefont
  {Martinis}}]{Galiautdinov2012}%
  \BibitemOpen
  \bibfield  {author} {\bibinfo {author} {\bibfnamefont {A.}~\bibnamefont
  {Galiautdinov}}, \bibinfo {author} {\bibfnamefont {A.~N.}\ \bibnamefont
  {Korotkov}},\ and\ \bibinfo {author} {\bibfnamefont {J.~M.}\ \bibnamefont
  {Martinis}},\ }\bibfield  {title} {\bibinfo {title} {Resonator--zero-qubit
  architecture for superconducting qubits},\ }\href
  {https://doi.org/10.1103/PhysRevA.85.042321} {\bibfield  {journal} {\bibinfo
  {journal} {Phys. Rev. A}\ }\textbf {\bibinfo {volume} {85}},\ \bibinfo
  {pages} {042321} (\bibinfo {year} {2012})}\BibitemShut {NoStop}%
\bibitem [{\citenamefont {Ghosh}\ \emph {et~al.}(2013)\citenamefont {Ghosh},
  \citenamefont {Galiautdinov}, \citenamefont {Zhou}, \citenamefont {Korotkov},
  \citenamefont {Martinis},\ and\ \citenamefont {Geller}}]{Ghosh2013}%
  \BibitemOpen
  \bibfield  {author} {\bibinfo {author} {\bibfnamefont {J.}~\bibnamefont
  {Ghosh}}, \bibinfo {author} {\bibfnamefont {A.}~\bibnamefont {Galiautdinov}},
  \bibinfo {author} {\bibfnamefont {Z.}~\bibnamefont {Zhou}}, \bibinfo {author}
  {\bibfnamefont {A.~N.}\ \bibnamefont {Korotkov}}, \bibinfo {author}
  {\bibfnamefont {J.~M.}\ \bibnamefont {Martinis}},\ and\ \bibinfo {author}
  {\bibfnamefont {M.~R.}\ \bibnamefont {Geller}},\ }\bibfield  {title}
  {\bibinfo {title} {High-fidelity controlled-${\ensuremath{\sigma}}^{Z}$ gate
  for resonator-based superconducting quantum computers},\ }\href
  {https://doi.org/10.1103/PhysRevA.87.022309} {\bibfield  {journal} {\bibinfo
  {journal} {Phys. Rev. A}\ }\textbf {\bibinfo {volume} {87}},\ \bibinfo
  {pages} {022309} (\bibinfo {year} {2013})}\BibitemShut {NoStop}%
\bibitem [{\citenamefont {Barends}\ \emph {et~al.}(2014)\citenamefont
  {Barends}, \citenamefont {Kelly}, \citenamefont {Megrant}, \citenamefont
  {Veitia}, \citenamefont {Sank}, \citenamefont {Jeffrey}, \citenamefont
  {White}, \citenamefont {Mutus}, \citenamefont {Fowler}, \citenamefont
  {Campbell}, \citenamefont {Chen}, \citenamefont {Chen}, \citenamefont
  {Chiaro}, \citenamefont {Dunsworth}, \citenamefont {Neill}, \citenamefont
  {O'Malley}, \citenamefont {Roushan}, \citenamefont {Vainsencher},
  \citenamefont {Wenner}, \citenamefont {Korotkov}, \citenamefont {Cleland},\
  and\ \citenamefont {Martinis}}]{Barends2014}%
  \BibitemOpen
  \bibfield  {author} {\bibinfo {author} {\bibfnamefont {R.}~\bibnamefont
  {Barends}}, \bibinfo {author} {\bibfnamefont {J.}~\bibnamefont {Kelly}},
  \bibinfo {author} {\bibfnamefont {A.}~\bibnamefont {Megrant}}, \bibinfo
  {author} {\bibfnamefont {A.}~\bibnamefont {Veitia}}, \bibinfo {author}
  {\bibfnamefont {D.}~\bibnamefont {Sank}}, \bibinfo {author} {\bibfnamefont
  {E.}~\bibnamefont {Jeffrey}}, \bibinfo {author} {\bibfnamefont {T.~C.}\
  \bibnamefont {White}}, \bibinfo {author} {\bibfnamefont {J.}~\bibnamefont
  {Mutus}}, \bibinfo {author} {\bibfnamefont {A.~G.}\ \bibnamefont {Fowler}},
  \bibinfo {author} {\bibfnamefont {B.}~\bibnamefont {Campbell}}, \bibinfo
  {author} {\bibfnamefont {Y.}~\bibnamefont {Chen}}, \bibinfo {author}
  {\bibfnamefont {Z.}~\bibnamefont {Chen}}, \bibinfo {author} {\bibfnamefont
  {B.}~\bibnamefont {Chiaro}}, \bibinfo {author} {\bibfnamefont
  {A.}~\bibnamefont {Dunsworth}}, \bibinfo {author} {\bibfnamefont
  {C.}~\bibnamefont {Neill}}, \bibinfo {author} {\bibfnamefont
  {P.}~\bibnamefont {O'Malley}}, \bibinfo {author} {\bibfnamefont
  {P.}~\bibnamefont {Roushan}}, \bibinfo {author} {\bibfnamefont
  {A.}~\bibnamefont {Vainsencher}}, \bibinfo {author} {\bibfnamefont
  {J.}~\bibnamefont {Wenner}}, \bibinfo {author} {\bibfnamefont {A.~N.}\
  \bibnamefont {Korotkov}}, \bibinfo {author} {\bibfnamefont {A.~N.}\
  \bibnamefont {Cleland}},\ and\ \bibinfo {author} {\bibfnamefont {J.~M.}\
  \bibnamefont {Martinis}},\ }\bibfield  {title} {\bibinfo {title}
  {Superconducting quantum circuits at the surface code threshold for fault
  tolerance},\ }\href {https://doi.org/10.1038/nature13171} {\bibfield
  {journal} {\bibinfo  {journal} {Nature}\ }\textbf {\bibinfo {volume} {508}},\
  \bibinfo {pages} {500} (\bibinfo {year} {2014})}\BibitemShut {NoStop}%
\bibitem [{\citenamefont {Kelly}\ \emph {et~al.}(2015)\citenamefont {Kelly},
  \citenamefont {Barends}, \citenamefont {Fowler}, \citenamefont {Megrant},
  \citenamefont {Jeffrey}, \citenamefont {White}, \citenamefont {Sank},
  \citenamefont {Mutus}, \citenamefont {Campbell}, \citenamefont {Chen},
  \citenamefont {Chen}, \citenamefont {Chiaro}, \citenamefont {Dunsworth},
  \citenamefont {Hoi}, \citenamefont {Neill}, \citenamefont {O'Malley},
  \citenamefont {Quintana}, \citenamefont {Roushan}, \citenamefont
  {Vainsencher}, \citenamefont {Wenner}, \citenamefont {Cleland},\ and\
  \citenamefont {Martinis}}]{Kelly2015}%
  \BibitemOpen
  \bibfield  {author} {\bibinfo {author} {\bibfnamefont {J.}~\bibnamefont
  {Kelly}}, \bibinfo {author} {\bibfnamefont {R.}~\bibnamefont {Barends}},
  \bibinfo {author} {\bibfnamefont {A.~G.}\ \bibnamefont {Fowler}}, \bibinfo
  {author} {\bibfnamefont {A.}~\bibnamefont {Megrant}}, \bibinfo {author}
  {\bibfnamefont {E.}~\bibnamefont {Jeffrey}}, \bibinfo {author} {\bibfnamefont
  {T.~C.}\ \bibnamefont {White}}, \bibinfo {author} {\bibfnamefont
  {D.}~\bibnamefont {Sank}}, \bibinfo {author} {\bibfnamefont {J.~Y.}\
  \bibnamefont {Mutus}}, \bibinfo {author} {\bibfnamefont {B.}~\bibnamefont
  {Campbell}}, \bibinfo {author} {\bibfnamefont {Y.}~\bibnamefont {Chen}},
  \bibinfo {author} {\bibfnamefont {Z.}~\bibnamefont {Chen}}, \bibinfo {author}
  {\bibfnamefont {B.}~\bibnamefont {Chiaro}}, \bibinfo {author} {\bibfnamefont
  {A.}~\bibnamefont {Dunsworth}}, \bibinfo {author} {\bibfnamefont {I.-C.}\
  \bibnamefont {Hoi}}, \bibinfo {author} {\bibfnamefont {C.}~\bibnamefont
  {Neill}}, \bibinfo {author} {\bibfnamefont {P.~J.~J.}\ \bibnamefont
  {O'Malley}}, \bibinfo {author} {\bibfnamefont {C.}~\bibnamefont {Quintana}},
  \bibinfo {author} {\bibfnamefont {P.}~\bibnamefont {Roushan}}, \bibinfo
  {author} {\bibfnamefont {A.}~\bibnamefont {Vainsencher}}, \bibinfo {author}
  {\bibfnamefont {J.}~\bibnamefont {Wenner}}, \bibinfo {author} {\bibfnamefont
  {A.~N.}\ \bibnamefont {Cleland}},\ and\ \bibinfo {author} {\bibfnamefont
  {J.~M.}\ \bibnamefont {Martinis}},\ }\bibfield  {title} {\bibinfo {title}
  {State preservation by repetitive error detection in a superconducting
  quantum circuit},\ }\href {https://doi.org/10.1038/nature14270} {\bibfield
  {journal} {\bibinfo  {journal} {Nature}\ }\textbf {\bibinfo {volume} {519}},\
  \bibinfo {pages} {66} (\bibinfo {year} {2015})}\BibitemShut {NoStop}%
\bibitem [{\citenamefont {Roushan}\ \emph {et~al.}(2017)\citenamefont
  {Roushan}, \citenamefont {Neill}, \citenamefont {Tangpanitanon},
  \citenamefont {Bastidas}, \citenamefont {Megrant}, \citenamefont {Barends},
  \citenamefont {Chen}, \citenamefont {Chen}, \citenamefont {Chiaro},
  \citenamefont {Dunsworth}, \citenamefont {Fowler}, \citenamefont {Foxen},
  \citenamefont {Giustina}, \citenamefont {Jeffrey}, \citenamefont {Kelly},
  \citenamefont {Lucero}, \citenamefont {Mutus}, \citenamefont {Neeley},
  \citenamefont {Quintana}, \citenamefont {Sank}, \citenamefont {Vainsencher},
  \citenamefont {Wenner}, \citenamefont {White}, \citenamefont {Neven},
  \citenamefont {Angelakis},\ and\ \citenamefont {Martinis}}]{Roushan2017}%
  \BibitemOpen
  \bibfield  {author} {\bibinfo {author} {\bibfnamefont {P.}~\bibnamefont
  {Roushan}}, \bibinfo {author} {\bibfnamefont {C.}~\bibnamefont {Neill}},
  \bibinfo {author} {\bibfnamefont {J.}~\bibnamefont {Tangpanitanon}}, \bibinfo
  {author} {\bibfnamefont {V.~M.}\ \bibnamefont {Bastidas}}, \bibinfo {author}
  {\bibfnamefont {A.}~\bibnamefont {Megrant}}, \bibinfo {author} {\bibfnamefont
  {R.}~\bibnamefont {Barends}}, \bibinfo {author} {\bibfnamefont
  {Y.}~\bibnamefont {Chen}}, \bibinfo {author} {\bibfnamefont {Z.}~\bibnamefont
  {Chen}}, \bibinfo {author} {\bibfnamefont {B.}~\bibnamefont {Chiaro}},
  \bibinfo {author} {\bibfnamefont {A.}~\bibnamefont {Dunsworth}}, \bibinfo
  {author} {\bibfnamefont {A.}~\bibnamefont {Fowler}}, \bibinfo {author}
  {\bibfnamefont {B.}~\bibnamefont {Foxen}}, \bibinfo {author} {\bibfnamefont
  {M.}~\bibnamefont {Giustina}}, \bibinfo {author} {\bibfnamefont
  {E.}~\bibnamefont {Jeffrey}}, \bibinfo {author} {\bibfnamefont
  {J.}~\bibnamefont {Kelly}}, \bibinfo {author} {\bibfnamefont
  {E.}~\bibnamefont {Lucero}}, \bibinfo {author} {\bibfnamefont
  {J.}~\bibnamefont {Mutus}}, \bibinfo {author} {\bibfnamefont
  {M.}~\bibnamefont {Neeley}}, \bibinfo {author} {\bibfnamefont
  {C.}~\bibnamefont {Quintana}}, \bibinfo {author} {\bibfnamefont
  {D.}~\bibnamefont {Sank}}, \bibinfo {author} {\bibfnamefont {A.}~\bibnamefont
  {Vainsencher}}, \bibinfo {author} {\bibfnamefont {J.}~\bibnamefont {Wenner}},
  \bibinfo {author} {\bibfnamefont {T.}~\bibnamefont {White}}, \bibinfo
  {author} {\bibfnamefont {H.}~\bibnamefont {Neven}}, \bibinfo {author}
  {\bibfnamefont {D.~G.}\ \bibnamefont {Angelakis}},\ and\ \bibinfo {author}
  {\bibfnamefont {J.}~\bibnamefont {Martinis}},\ }\bibfield  {title} {\bibinfo
  {title} {Spectroscopic signatures of localization with interacting photons in
  superconducting qubits},\ }\href {https://doi.org/10.1126/science.aao1401}
  {\bibfield  {journal} {\bibinfo  {journal} {Science}\ }\textbf {\bibinfo
  {volume} {358}},\ \bibinfo {pages} {1175} (\bibinfo {year}
  {2017})}\BibitemShut {NoStop}%
\bibitem [{\citenamefont {Lucero}\ \emph {et~al.}(2008)\citenamefont {Lucero},
  \citenamefont {Hofheinz}, \citenamefont {Ansmann}, \citenamefont {Bialczak},
  \citenamefont {Katz}, \citenamefont {Neeley}, \citenamefont {O'Connell},
  \citenamefont {Wang}, \citenamefont {Cleland},\ and\ \citenamefont
  {Martinis}}]{Lucero2008}%
  \BibitemOpen
  \bibfield  {author} {\bibinfo {author} {\bibfnamefont {E.}~\bibnamefont
  {Lucero}}, \bibinfo {author} {\bibfnamefont {M.}~\bibnamefont {Hofheinz}},
  \bibinfo {author} {\bibfnamefont {M.}~\bibnamefont {Ansmann}}, \bibinfo
  {author} {\bibfnamefont {R.~C.}\ \bibnamefont {Bialczak}}, \bibinfo {author}
  {\bibfnamefont {N.}~\bibnamefont {Katz}}, \bibinfo {author} {\bibfnamefont
  {M.}~\bibnamefont {Neeley}}, \bibinfo {author} {\bibfnamefont {A.~D.}\
  \bibnamefont {O'Connell}}, \bibinfo {author} {\bibfnamefont {H.}~\bibnamefont
  {Wang}}, \bibinfo {author} {\bibfnamefont {A.~N.}\ \bibnamefont {Cleland}},\
  and\ \bibinfo {author} {\bibfnamefont {J.~M.}\ \bibnamefont {Martinis}},\
  }\bibfield  {title} {\bibinfo {title} {High-fidelity gates in a single
  josephson qubit},\ }\href {https://doi.org/10.1103/PhysRevLett.100.247001}
  {\bibfield  {journal} {\bibinfo  {journal} {Phys. Rev. Lett.}\ }\textbf
  {\bibinfo {volume} {100}},\ \bibinfo {pages} {247001} (\bibinfo {year}
  {2008})}\BibitemShut {NoStop}%
\bibitem [{\citenamefont {Kelly}\ \emph {et~al.}(2014)\citenamefont {Kelly},
  \citenamefont {Barends}, \citenamefont {Campbell}, \citenamefont {Chen},
  \citenamefont {Chen}, \citenamefont {Chiaro}, \citenamefont {Dunsworth},
  \citenamefont {Fowler}, \citenamefont {Hoi}, \citenamefont {Jeffrey},
  \citenamefont {Megrant}, \citenamefont {Mutus}, \citenamefont {Neill},
  \citenamefont {O'Malley}, \citenamefont {Quintana}, \citenamefont {Roushan},
  \citenamefont {Sank}, \citenamefont {Vainsencher}, \citenamefont {Wenner},
  \citenamefont {White}, \citenamefont {Cleland},\ and\ \citenamefont
  {Martinis}}]{Kelly2014}%
  \BibitemOpen
  \bibfield  {author} {\bibinfo {author} {\bibfnamefont {J.}~\bibnamefont
  {Kelly}}, \bibinfo {author} {\bibfnamefont {R.}~\bibnamefont {Barends}},
  \bibinfo {author} {\bibfnamefont {B.}~\bibnamefont {Campbell}}, \bibinfo
  {author} {\bibfnamefont {Y.}~\bibnamefont {Chen}}, \bibinfo {author}
  {\bibfnamefont {Z.}~\bibnamefont {Chen}}, \bibinfo {author} {\bibfnamefont
  {B.}~\bibnamefont {Chiaro}}, \bibinfo {author} {\bibfnamefont
  {A.}~\bibnamefont {Dunsworth}}, \bibinfo {author} {\bibfnamefont {A.~G.}\
  \bibnamefont {Fowler}}, \bibinfo {author} {\bibfnamefont {I.-C.}\
  \bibnamefont {Hoi}}, \bibinfo {author} {\bibfnamefont {E.}~\bibnamefont
  {Jeffrey}}, \bibinfo {author} {\bibfnamefont {A.}~\bibnamefont {Megrant}},
  \bibinfo {author} {\bibfnamefont {J.}~\bibnamefont {Mutus}}, \bibinfo
  {author} {\bibfnamefont {C.}~\bibnamefont {Neill}}, \bibinfo {author}
  {\bibfnamefont {P.~J.~J.}\ \bibnamefont {O'Malley}}, \bibinfo {author}
  {\bibfnamefont {C.}~\bibnamefont {Quintana}}, \bibinfo {author}
  {\bibfnamefont {P.}~\bibnamefont {Roushan}}, \bibinfo {author} {\bibfnamefont
  {D.}~\bibnamefont {Sank}}, \bibinfo {author} {\bibfnamefont {A.}~\bibnamefont
  {Vainsencher}}, \bibinfo {author} {\bibfnamefont {J.}~\bibnamefont {Wenner}},
  \bibinfo {author} {\bibfnamefont {T.~C.}\ \bibnamefont {White}}, \bibinfo
  {author} {\bibfnamefont {A.~N.}\ \bibnamefont {Cleland}},\ and\ \bibinfo
  {author} {\bibfnamefont {J.~M.}\ \bibnamefont {Martinis}},\ }\bibfield
  {title} {\bibinfo {title} {Optimal quantum control using randomized
  benchmarking},\ }\href {https://doi.org/10.1103/PhysRevLett.112.240504}
  {\bibfield  {journal} {\bibinfo  {journal} {Phys. Rev. Lett.}\ }\textbf
  {\bibinfo {volume} {112}},\ \bibinfo {pages} {240504} (\bibinfo {year}
  {2014})}\BibitemShut {NoStop}%
\bibitem [{\citenamefont {Arute}\ \emph {et~al.}(2019)\citenamefont {Arute},
  \citenamefont {Arya}, \citenamefont {Babbush}, \citenamefont {Bacon},
  \citenamefont {Bardin}, \citenamefont {Barends}, \citenamefont {Biswas},
  \citenamefont {Boixo}, \citenamefont {Brandao}, \citenamefont {Buell},
  \citenamefont {Burkett}, \citenamefont {Chen}, \citenamefont {Chen},
  \citenamefont {Chiaro}, \citenamefont {Collins}, \citenamefont {Courtney},
  \citenamefont {Dunsworth}, \citenamefont {Farhi}, \citenamefont {Foxen},
  \citenamefont {Fowler}, \citenamefont {Gidney}, \citenamefont {Giustina},
  \citenamefont {Graff}, \citenamefont {Guerin}, \citenamefont {Habegger},
  \citenamefont {Harrigan}, \citenamefont {Hartmann}, \citenamefont {Ho},
  \citenamefont {Hoffmann}, \citenamefont {Huang}, \citenamefont {Humble},
  \citenamefont {Isakov}, \citenamefont {Jeffrey}, \citenamefont {Jiang},
  \citenamefont {Kafri}, \citenamefont {Kechedzhi}, \citenamefont {Kelly},
  \citenamefont {Klimov}, \citenamefont {Knysh}, \citenamefont {Korotkov},
  \citenamefont {Kostritsa}, \citenamefont {Landhuis}, \citenamefont
  {Lindmark}, \citenamefont {Lucero}, \citenamefont {Lyakh}, \citenamefont
  {Mandrà}, \citenamefont {McClean}, \citenamefont {McEwen}, \citenamefont
  {Megrant}, \citenamefont {Mi}, \citenamefont {Michielsen}, \citenamefont
  {Mohseni}, \citenamefont {Mutus}, \citenamefont {Naaman}, \citenamefont
  {Neeley}, \citenamefont {Neill}, \citenamefont {Niu}, \citenamefont {Ostby},
  \citenamefont {Petukhov}, \citenamefont {Platt}, \citenamefont {Quintana},
  \citenamefont {Rieffel}, \citenamefont {Roushan}, \citenamefont {Rubin},
  \citenamefont {Sank}, \citenamefont {Satzinger}, \citenamefont {Smelyanskiy},
  \citenamefont {Sung}, \citenamefont {Trevithick}, \citenamefont
  {Vainsencher}, \citenamefont {Villalonga}, \citenamefont {White},
  \citenamefont {Yao}, \citenamefont {Yeh}, \citenamefont {Zalcman},
  \citenamefont {Neven},\ and\ \citenamefont {Martinis}}]{Arute2019}%
  \BibitemOpen
  \bibfield  {author} {\bibinfo {author} {\bibfnamefont {F.}~\bibnamefont
  {Arute}}, \bibinfo {author} {\bibfnamefont {K.}~\bibnamefont {Arya}},
  \bibinfo {author} {\bibfnamefont {R.}~\bibnamefont {Babbush}}, \bibinfo
  {author} {\bibfnamefont {D.}~\bibnamefont {Bacon}}, \bibinfo {author}
  {\bibfnamefont {J.~C.}\ \bibnamefont {Bardin}}, \bibinfo {author}
  {\bibfnamefont {R.}~\bibnamefont {Barends}}, \bibinfo {author} {\bibfnamefont
  {R.}~\bibnamefont {Biswas}}, \bibinfo {author} {\bibfnamefont
  {S.}~\bibnamefont {Boixo}}, \bibinfo {author} {\bibfnamefont {F.~G. S.~L.}\
  \bibnamefont {Brandao}}, \bibinfo {author} {\bibfnamefont {D.~A.}\
  \bibnamefont {Buell}}, \bibinfo {author} {\bibfnamefont {B.}~\bibnamefont
  {Burkett}}, \bibinfo {author} {\bibfnamefont {Y.}~\bibnamefont {Chen}},
  \bibinfo {author} {\bibfnamefont {Z.}~\bibnamefont {Chen}}, \bibinfo {author}
  {\bibfnamefont {B.}~\bibnamefont {Chiaro}}, \bibinfo {author} {\bibfnamefont
  {R.}~\bibnamefont {Collins}}, \bibinfo {author} {\bibfnamefont
  {W.}~\bibnamefont {Courtney}}, \bibinfo {author} {\bibfnamefont
  {A.}~\bibnamefont {Dunsworth}}, \bibinfo {author} {\bibfnamefont
  {E.}~\bibnamefont {Farhi}}, \bibinfo {author} {\bibfnamefont
  {B.}~\bibnamefont {Foxen}}, \bibinfo {author} {\bibfnamefont
  {A.}~\bibnamefont {Fowler}}, \bibinfo {author} {\bibfnamefont
  {C.}~\bibnamefont {Gidney}}, \bibinfo {author} {\bibfnamefont
  {M.}~\bibnamefont {Giustina}}, \bibinfo {author} {\bibfnamefont
  {R.}~\bibnamefont {Graff}}, \bibinfo {author} {\bibfnamefont
  {K.}~\bibnamefont {Guerin}}, \bibinfo {author} {\bibfnamefont
  {S.}~\bibnamefont {Habegger}}, \bibinfo {author} {\bibfnamefont {M.~P.}\
  \bibnamefont {Harrigan}}, \bibinfo {author} {\bibfnamefont {M.~J.}\
  \bibnamefont {Hartmann}}, \bibinfo {author} {\bibfnamefont {A.}~\bibnamefont
  {Ho}}, \bibinfo {author} {\bibfnamefont {M.}~\bibnamefont {Hoffmann}},
  \bibinfo {author} {\bibfnamefont {T.}~\bibnamefont {Huang}}, \bibinfo
  {author} {\bibfnamefont {T.~S.}\ \bibnamefont {Humble}}, \bibinfo {author}
  {\bibfnamefont {S.~V.}\ \bibnamefont {Isakov}}, \bibinfo {author}
  {\bibfnamefont {E.}~\bibnamefont {Jeffrey}}, \bibinfo {author} {\bibfnamefont
  {Z.}~\bibnamefont {Jiang}}, \bibinfo {author} {\bibfnamefont
  {D.}~\bibnamefont {Kafri}}, \bibinfo {author} {\bibfnamefont
  {K.}~\bibnamefont {Kechedzhi}}, \bibinfo {author} {\bibfnamefont
  {J.}~\bibnamefont {Kelly}}, \bibinfo {author} {\bibfnamefont {P.~V.}\
  \bibnamefont {Klimov}}, \bibinfo {author} {\bibfnamefont {S.}~\bibnamefont
  {Knysh}}, \bibinfo {author} {\bibfnamefont {A.}~\bibnamefont {Korotkov}},
  \bibinfo {author} {\bibfnamefont {F.}~\bibnamefont {Kostritsa}}, \bibinfo
  {author} {\bibfnamefont {D.}~\bibnamefont {Landhuis}}, \bibinfo {author}
  {\bibfnamefont {M.}~\bibnamefont {Lindmark}}, \bibinfo {author}
  {\bibfnamefont {E.}~\bibnamefont {Lucero}}, \bibinfo {author} {\bibfnamefont
  {D.}~\bibnamefont {Lyakh}}, \bibinfo {author} {\bibfnamefont
  {S.}~\bibnamefont {Mandrà}}, \bibinfo {author} {\bibfnamefont {J.~R.}\
  \bibnamefont {McClean}}, \bibinfo {author} {\bibfnamefont {M.}~\bibnamefont
  {McEwen}}, \bibinfo {author} {\bibfnamefont {A.}~\bibnamefont {Megrant}},
  \bibinfo {author} {\bibfnamefont {X.}~\bibnamefont {Mi}}, \bibinfo {author}
  {\bibfnamefont {K.}~\bibnamefont {Michielsen}}, \bibinfo {author}
  {\bibfnamefont {M.}~\bibnamefont {Mohseni}}, \bibinfo {author} {\bibfnamefont
  {J.}~\bibnamefont {Mutus}}, \bibinfo {author} {\bibfnamefont
  {O.}~\bibnamefont {Naaman}}, \bibinfo {author} {\bibfnamefont
  {M.}~\bibnamefont {Neeley}}, \bibinfo {author} {\bibfnamefont
  {C.}~\bibnamefont {Neill}}, \bibinfo {author} {\bibfnamefont {M.~Y.}\
  \bibnamefont {Niu}}, \bibinfo {author} {\bibfnamefont {E.}~\bibnamefont
  {Ostby}}, \bibinfo {author} {\bibfnamefont {A.}~\bibnamefont {Petukhov}},
  \bibinfo {author} {\bibfnamefont {J.~C.}\ \bibnamefont {Platt}}, \bibinfo
  {author} {\bibfnamefont {C.}~\bibnamefont {Quintana}}, \bibinfo {author}
  {\bibfnamefont {E.~G.}\ \bibnamefont {Rieffel}}, \bibinfo {author}
  {\bibfnamefont {P.}~\bibnamefont {Roushan}}, \bibinfo {author} {\bibfnamefont
  {N.~C.}\ \bibnamefont {Rubin}}, \bibinfo {author} {\bibfnamefont
  {D.}~\bibnamefont {Sank}}, \bibinfo {author} {\bibfnamefont {K.~J.}\
  \bibnamefont {Satzinger}}, \bibinfo {author} {\bibfnamefont {V.}~\bibnamefont
  {Smelyanskiy}}, \bibinfo {author} {\bibfnamefont {K.~J.}\ \bibnamefont
  {Sung}}, \bibinfo {author} {\bibfnamefont {M.~D.}\ \bibnamefont
  {Trevithick}}, \bibinfo {author} {\bibfnamefont {A.}~\bibnamefont
  {Vainsencher}}, \bibinfo {author} {\bibfnamefont {B.}~\bibnamefont
  {Villalonga}}, \bibinfo {author} {\bibfnamefont {T.}~\bibnamefont {White}},
  \bibinfo {author} {\bibfnamefont {Z.~J.}\ \bibnamefont {Yao}}, \bibinfo
  {author} {\bibfnamefont {P.}~\bibnamefont {Yeh}}, \bibinfo {author}
  {\bibfnamefont {A.}~\bibnamefont {Zalcman}}, \bibinfo {author} {\bibfnamefont
  {H.}~\bibnamefont {Neven}},\ and\ \bibinfo {author} {\bibfnamefont {J.~M.}\
  \bibnamefont {Martinis}},\ }\bibfield  {title} {\bibinfo {title} {{Quantum
  supremacy using a programmable superconducting processor}},\ }\href
  {https://doi.org/10.1038/s41586-019-1666-5} {\bibfield  {journal} {\bibinfo
  {journal} {Nature}\ }\textbf {\bibinfo {volume} {574}},\ \bibinfo {pages}
  {505} (\bibinfo {year} {2019})}\BibitemShut {NoStop}%
\bibitem [{\citenamefont {Foxen}\ \emph {et~al.}(2020)\citenamefont {Foxen},
  \citenamefont {Neill}, \citenamefont {Dunsworth}, \citenamefont {Roushan},
  \citenamefont {Chiaro}, \citenamefont {Megrant}, \citenamefont {Kelly},
  \citenamefont {Chen}, \citenamefont {Satzinger}, \citenamefont {Barends},
  \citenamefont {Arute}, \citenamefont {Arya}, \citenamefont {Babbush},
  \citenamefont {Bacon}, \citenamefont {Bardin}, \citenamefont {Boixo},
  \citenamefont {Buell}, \citenamefont {Burkett}, \citenamefont {Chen},
  \citenamefont {Collins}, \citenamefont {Farhi}, \citenamefont {Fowler},
  \citenamefont {Gidney}, \citenamefont {Giustina}, \citenamefont {Graff},
  \citenamefont {Harrigan}, \citenamefont {Huang}, \citenamefont {Isakov},
  \citenamefont {Jeffrey}, \citenamefont {Jiang}, \citenamefont {Kafri},
  \citenamefont {Kechedzhi}, \citenamefont {Klimov}, \citenamefont {Korotkov},
  \citenamefont {Kostritsa}, \citenamefont {Landhuis}, \citenamefont {Lucero},
  \citenamefont {McClean}, \citenamefont {McEwen}, \citenamefont {Mi},
  \citenamefont {Mohseni}, \citenamefont {Mutus}, \citenamefont {Naaman},
  \citenamefont {Neeley}, \citenamefont {Niu}, \citenamefont {Petukhov},
  \citenamefont {Quintana}, \citenamefont {Rubin}, \citenamefont {Sank},
  \citenamefont {Smelyanskiy}, \citenamefont {Vainsencher}, \citenamefont
  {White}, \citenamefont {Yao}, \citenamefont {Yeh}, \citenamefont {Zalcman},
  \citenamefont {Neven},\ and\ \citenamefont {Martinis}}]{Foxen2020}%
  \BibitemOpen
  \bibfield  {author} {\bibinfo {author} {\bibfnamefont {B.}~\bibnamefont
  {Foxen}}, \bibinfo {author} {\bibfnamefont {C.}~\bibnamefont {Neill}},
  \bibinfo {author} {\bibfnamefont {A.}~\bibnamefont {Dunsworth}}, \bibinfo
  {author} {\bibfnamefont {P.}~\bibnamefont {Roushan}}, \bibinfo {author}
  {\bibfnamefont {B.}~\bibnamefont {Chiaro}}, \bibinfo {author} {\bibfnamefont
  {A.}~\bibnamefont {Megrant}}, \bibinfo {author} {\bibfnamefont
  {J.}~\bibnamefont {Kelly}}, \bibinfo {author} {\bibfnamefont
  {Z.}~\bibnamefont {Chen}}, \bibinfo {author} {\bibfnamefont {K.}~\bibnamefont
  {Satzinger}}, \bibinfo {author} {\bibfnamefont {R.}~\bibnamefont {Barends}},
  \bibinfo {author} {\bibfnamefont {F.}~\bibnamefont {Arute}}, \bibinfo
  {author} {\bibfnamefont {K.}~\bibnamefont {Arya}}, \bibinfo {author}
  {\bibfnamefont {R.}~\bibnamefont {Babbush}}, \bibinfo {author} {\bibfnamefont
  {D.}~\bibnamefont {Bacon}}, \bibinfo {author} {\bibfnamefont {J.~C.}\
  \bibnamefont {Bardin}}, \bibinfo {author} {\bibfnamefont {S.}~\bibnamefont
  {Boixo}}, \bibinfo {author} {\bibfnamefont {D.}~\bibnamefont {Buell}},
  \bibinfo {author} {\bibfnamefont {B.}~\bibnamefont {Burkett}}, \bibinfo
  {author} {\bibfnamefont {Y.}~\bibnamefont {Chen}}, \bibinfo {author}
  {\bibfnamefont {R.}~\bibnamefont {Collins}}, \bibinfo {author} {\bibfnamefont
  {E.}~\bibnamefont {Farhi}}, \bibinfo {author} {\bibfnamefont
  {A.}~\bibnamefont {Fowler}}, \bibinfo {author} {\bibfnamefont
  {C.}~\bibnamefont {Gidney}}, \bibinfo {author} {\bibfnamefont
  {M.}~\bibnamefont {Giustina}}, \bibinfo {author} {\bibfnamefont
  {R.}~\bibnamefont {Graff}}, \bibinfo {author} {\bibfnamefont
  {M.}~\bibnamefont {Harrigan}}, \bibinfo {author} {\bibfnamefont
  {T.}~\bibnamefont {Huang}}, \bibinfo {author} {\bibfnamefont {S.~V.}\
  \bibnamefont {Isakov}}, \bibinfo {author} {\bibfnamefont {E.}~\bibnamefont
  {Jeffrey}}, \bibinfo {author} {\bibfnamefont {Z.}~\bibnamefont {Jiang}},
  \bibinfo {author} {\bibfnamefont {D.}~\bibnamefont {Kafri}}, \bibinfo
  {author} {\bibfnamefont {K.}~\bibnamefont {Kechedzhi}}, \bibinfo {author}
  {\bibfnamefont {P.}~\bibnamefont {Klimov}}, \bibinfo {author} {\bibfnamefont
  {A.}~\bibnamefont {Korotkov}}, \bibinfo {author} {\bibfnamefont
  {F.}~\bibnamefont {Kostritsa}}, \bibinfo {author} {\bibfnamefont
  {D.}~\bibnamefont {Landhuis}}, \bibinfo {author} {\bibfnamefont
  {E.}~\bibnamefont {Lucero}}, \bibinfo {author} {\bibfnamefont
  {J.}~\bibnamefont {McClean}}, \bibinfo {author} {\bibfnamefont
  {M.}~\bibnamefont {McEwen}}, \bibinfo {author} {\bibfnamefont
  {X.}~\bibnamefont {Mi}}, \bibinfo {author} {\bibfnamefont {M.}~\bibnamefont
  {Mohseni}}, \bibinfo {author} {\bibfnamefont {J.~Y.}\ \bibnamefont {Mutus}},
  \bibinfo {author} {\bibfnamefont {O.}~\bibnamefont {Naaman}}, \bibinfo
  {author} {\bibfnamefont {M.}~\bibnamefont {Neeley}}, \bibinfo {author}
  {\bibfnamefont {M.}~\bibnamefont {Niu}}, \bibinfo {author} {\bibfnamefont
  {A.}~\bibnamefont {Petukhov}}, \bibinfo {author} {\bibfnamefont
  {C.}~\bibnamefont {Quintana}}, \bibinfo {author} {\bibfnamefont
  {N.}~\bibnamefont {Rubin}}, \bibinfo {author} {\bibfnamefont
  {D.}~\bibnamefont {Sank}}, \bibinfo {author} {\bibfnamefont {V.}~\bibnamefont
  {Smelyanskiy}}, \bibinfo {author} {\bibfnamefont {A.}~\bibnamefont
  {Vainsencher}}, \bibinfo {author} {\bibfnamefont {T.~C.}\ \bibnamefont
  {White}}, \bibinfo {author} {\bibfnamefont {Z.}~\bibnamefont {Yao}}, \bibinfo
  {author} {\bibfnamefont {P.}~\bibnamefont {Yeh}}, \bibinfo {author}
  {\bibfnamefont {A.}~\bibnamefont {Zalcman}}, \bibinfo {author} {\bibfnamefont
  {H.}~\bibnamefont {Neven}},\ and\ \bibinfo {author} {\bibfnamefont {J.~M.}\
  \bibnamefont {Martinis}} (\bibinfo {collaboration} {Google AI Quantum}),\
  }\bibfield  {title} {\bibinfo {title} {Demonstrating a continuous set of
  two-qubit gates for near-term quantum algorithms},\ }\href
  {https://doi.org/10.1103/PhysRevLett.125.120504} {\bibfield  {journal}
  {\bibinfo  {journal} {Phys. Rev. Lett.}\ }\textbf {\bibinfo {volume} {125}},\
  \bibinfo {pages} {120504} (\bibinfo {year} {2020})}\BibitemShut {NoStop}%
\bibitem [{\citenamefont {Harrigan}\ \emph {et~al.}(2021)\citenamefont
  {Harrigan}, \citenamefont {Sung}, \citenamefont {Neeley}, \citenamefont
  {Satzinger}, \citenamefont {Arute}, \citenamefont {Arya}, \citenamefont
  {Atalaya}, \citenamefont {Bardin}, \citenamefont {Barends}, \citenamefont
  {Boixo}, \citenamefont {Broughton}, \citenamefont {Buckley}, \citenamefont
  {Buell}, \citenamefont {Burkett}, \citenamefont {Bushnell}, \citenamefont
  {Chen}, \citenamefont {Chen}, \citenamefont {Chiaro}, \citenamefont
  {Collins}, \citenamefont {Courtney}, \citenamefont {Demura}, \citenamefont
  {Dunsworth}, \citenamefont {Eppens}, \citenamefont {Fowler}, \citenamefont
  {Foxen}, \citenamefont {Gidney}, \citenamefont {Giustina}, \citenamefont
  {Graff}, \citenamefont {Habegger}, \citenamefont {Ho}, \citenamefont {Hong},
  \citenamefont {Huang}, \citenamefont {Ioffe}, \citenamefont {Isakov},
  \citenamefont {Jeffrey}, \citenamefont {Jiang}, \citenamefont {Jones},
  \citenamefont {Kafri}, \citenamefont {Kechedzhi}, \citenamefont {Kelly},
  \citenamefont {Kim}, \citenamefont {Klimov}, \citenamefont {Korotkov},
  \citenamefont {Kostritsa}, \citenamefont {Landhuis}, \citenamefont {Laptev},
  \citenamefont {Lindmark}, \citenamefont {Leib}, \citenamefont {Martin},
  \citenamefont {Martinis}, \citenamefont {McClean}, \citenamefont {McEwen},
  \citenamefont {Megrant}, \citenamefont {Mi}, \citenamefont {Mohseni},
  \citenamefont {Mruczkiewicz}, \citenamefont {Mutus}, \citenamefont {Naaman},
  \citenamefont {Neill}, \citenamefont {Neukart}, \citenamefont {Niu},
  \citenamefont {O'Brien}, \citenamefont {O'Gorman}, \citenamefont {Ostby},
  \citenamefont {Petukhov}, \citenamefont {Putterman}, \citenamefont
  {Quintana}, \citenamefont {Roushan}, \citenamefont {Rubin}, \citenamefont
  {Sank}, \citenamefont {Skolik}, \citenamefont {Smelyanskiy}, \citenamefont
  {Strain}, \citenamefont {Streif}, \citenamefont {Szalay}, \citenamefont
  {Vainsencher}, \citenamefont {White}, \citenamefont {Yao}, \citenamefont
  {Yeh}, \citenamefont {Zalcman}, \citenamefont {Zhou}, \citenamefont {Neven},
  \citenamefont {Bacon}, \citenamefont {Lucero}, \citenamefont {Farhi},\ and\
  \citenamefont {Babbush}}]{Harrigan2021}%
  \BibitemOpen
  \bibfield  {author} {\bibinfo {author} {\bibfnamefont {M.~P.}\ \bibnamefont
  {Harrigan}}, \bibinfo {author} {\bibfnamefont {K.~J.}\ \bibnamefont {Sung}},
  \bibinfo {author} {\bibfnamefont {M.}~\bibnamefont {Neeley}}, \bibinfo
  {author} {\bibfnamefont {K.~J.}\ \bibnamefont {Satzinger}}, \bibinfo {author}
  {\bibfnamefont {F.}~\bibnamefont {Arute}}, \bibinfo {author} {\bibfnamefont
  {K.}~\bibnamefont {Arya}}, \bibinfo {author} {\bibfnamefont {J.}~\bibnamefont
  {Atalaya}}, \bibinfo {author} {\bibfnamefont {J.~C.}\ \bibnamefont {Bardin}},
  \bibinfo {author} {\bibfnamefont {R.}~\bibnamefont {Barends}}, \bibinfo
  {author} {\bibfnamefont {S.}~\bibnamefont {Boixo}}, \bibinfo {author}
  {\bibfnamefont {M.}~\bibnamefont {Broughton}}, \bibinfo {author}
  {\bibfnamefont {B.~B.}\ \bibnamefont {Buckley}}, \bibinfo {author}
  {\bibfnamefont {D.~A.}\ \bibnamefont {Buell}}, \bibinfo {author}
  {\bibfnamefont {B.}~\bibnamefont {Burkett}}, \bibinfo {author} {\bibfnamefont
  {N.}~\bibnamefont {Bushnell}}, \bibinfo {author} {\bibfnamefont
  {Y.}~\bibnamefont {Chen}}, \bibinfo {author} {\bibfnamefont {Z.}~\bibnamefont
  {Chen}}, \bibinfo {author} {\bibfnamefont {B.}~\bibnamefont {Chiaro}},
  \bibinfo {author} {\bibfnamefont {R.}~\bibnamefont {Collins}}, \bibinfo
  {author} {\bibfnamefont {W.}~\bibnamefont {Courtney}}, \bibinfo {author}
  {\bibfnamefont {S.}~\bibnamefont {Demura}}, \bibinfo {author} {\bibfnamefont
  {A.}~\bibnamefont {Dunsworth}}, \bibinfo {author} {\bibfnamefont
  {D.}~\bibnamefont {Eppens}}, \bibinfo {author} {\bibfnamefont
  {A.}~\bibnamefont {Fowler}}, \bibinfo {author} {\bibfnamefont
  {B.}~\bibnamefont {Foxen}}, \bibinfo {author} {\bibfnamefont
  {C.}~\bibnamefont {Gidney}}, \bibinfo {author} {\bibfnamefont
  {M.}~\bibnamefont {Giustina}}, \bibinfo {author} {\bibfnamefont
  {R.}~\bibnamefont {Graff}}, \bibinfo {author} {\bibfnamefont
  {S.}~\bibnamefont {Habegger}}, \bibinfo {author} {\bibfnamefont
  {A.}~\bibnamefont {Ho}}, \bibinfo {author} {\bibfnamefont {S.}~\bibnamefont
  {Hong}}, \bibinfo {author} {\bibfnamefont {T.}~\bibnamefont {Huang}},
  \bibinfo {author} {\bibfnamefont {L.~B.}\ \bibnamefont {Ioffe}}, \bibinfo
  {author} {\bibfnamefont {S.~V.}\ \bibnamefont {Isakov}}, \bibinfo {author}
  {\bibfnamefont {E.}~\bibnamefont {Jeffrey}}, \bibinfo {author} {\bibfnamefont
  {Z.}~\bibnamefont {Jiang}}, \bibinfo {author} {\bibfnamefont
  {C.}~\bibnamefont {Jones}}, \bibinfo {author} {\bibfnamefont
  {D.}~\bibnamefont {Kafri}}, \bibinfo {author} {\bibfnamefont
  {K.}~\bibnamefont {Kechedzhi}}, \bibinfo {author} {\bibfnamefont
  {J.}~\bibnamefont {Kelly}}, \bibinfo {author} {\bibfnamefont
  {S.}~\bibnamefont {Kim}}, \bibinfo {author} {\bibfnamefont {P.~V.}\
  \bibnamefont {Klimov}}, \bibinfo {author} {\bibfnamefont {A.~N.}\
  \bibnamefont {Korotkov}}, \bibinfo {author} {\bibfnamefont {F.}~\bibnamefont
  {Kostritsa}}, \bibinfo {author} {\bibfnamefont {D.}~\bibnamefont {Landhuis}},
  \bibinfo {author} {\bibfnamefont {P.}~\bibnamefont {Laptev}}, \bibinfo
  {author} {\bibfnamefont {M.}~\bibnamefont {Lindmark}}, \bibinfo {author}
  {\bibfnamefont {M.}~\bibnamefont {Leib}}, \bibinfo {author} {\bibfnamefont
  {O.}~\bibnamefont {Martin}}, \bibinfo {author} {\bibfnamefont {J.~M.}\
  \bibnamefont {Martinis}}, \bibinfo {author} {\bibfnamefont {J.~R.}\
  \bibnamefont {McClean}}, \bibinfo {author} {\bibfnamefont {M.}~\bibnamefont
  {McEwen}}, \bibinfo {author} {\bibfnamefont {A.}~\bibnamefont {Megrant}},
  \bibinfo {author} {\bibfnamefont {X.}~\bibnamefont {Mi}}, \bibinfo {author}
  {\bibfnamefont {M.}~\bibnamefont {Mohseni}}, \bibinfo {author} {\bibfnamefont
  {W.}~\bibnamefont {Mruczkiewicz}}, \bibinfo {author} {\bibfnamefont
  {J.}~\bibnamefont {Mutus}}, \bibinfo {author} {\bibfnamefont
  {O.}~\bibnamefont {Naaman}}, \bibinfo {author} {\bibfnamefont
  {C.}~\bibnamefont {Neill}}, \bibinfo {author} {\bibfnamefont
  {F.}~\bibnamefont {Neukart}}, \bibinfo {author} {\bibfnamefont {M.~Y.}\
  \bibnamefont {Niu}}, \bibinfo {author} {\bibfnamefont {T.~E.}\ \bibnamefont
  {O'Brien}}, \bibinfo {author} {\bibfnamefont {B.}~\bibnamefont {O'Gorman}},
  \bibinfo {author} {\bibfnamefont {E.}~\bibnamefont {Ostby}}, \bibinfo
  {author} {\bibfnamefont {A.}~\bibnamefont {Petukhov}}, \bibinfo {author}
  {\bibfnamefont {H.}~\bibnamefont {Putterman}}, \bibinfo {author}
  {\bibfnamefont {C.}~\bibnamefont {Quintana}}, \bibinfo {author}
  {\bibfnamefont {P.}~\bibnamefont {Roushan}}, \bibinfo {author} {\bibfnamefont
  {N.~C.}\ \bibnamefont {Rubin}}, \bibinfo {author} {\bibfnamefont
  {D.}~\bibnamefont {Sank}}, \bibinfo {author} {\bibfnamefont {A.}~\bibnamefont
  {Skolik}}, \bibinfo {author} {\bibfnamefont {V.}~\bibnamefont {Smelyanskiy}},
  \bibinfo {author} {\bibfnamefont {D.}~\bibnamefont {Strain}}, \bibinfo
  {author} {\bibfnamefont {M.}~\bibnamefont {Streif}}, \bibinfo {author}
  {\bibfnamefont {M.}~\bibnamefont {Szalay}}, \bibinfo {author} {\bibfnamefont
  {A.}~\bibnamefont {Vainsencher}}, \bibinfo {author} {\bibfnamefont
  {T.}~\bibnamefont {White}}, \bibinfo {author} {\bibfnamefont {Z.~J.}\
  \bibnamefont {Yao}}, \bibinfo {author} {\bibfnamefont {P.}~\bibnamefont
  {Yeh}}, \bibinfo {author} {\bibfnamefont {A.}~\bibnamefont {Zalcman}},
  \bibinfo {author} {\bibfnamefont {L.}~\bibnamefont {Zhou}}, \bibinfo {author}
  {\bibfnamefont {H.}~\bibnamefont {Neven}}, \bibinfo {author} {\bibfnamefont
  {D.}~\bibnamefont {Bacon}}, \bibinfo {author} {\bibfnamefont
  {E.}~\bibnamefont {Lucero}}, \bibinfo {author} {\bibfnamefont
  {E.}~\bibnamefont {Farhi}},\ and\ \bibinfo {author} {\bibfnamefont
  {R.}~\bibnamefont {Babbush}},\ }\bibfield  {title} {\bibinfo {title} {Quantum
  approximate optimization of non-planar graph problems on a planar
  superconducting processor},\ }\href
  {https://doi.org/10.1038/s41567-020-01105-y} {\bibfield  {journal} {\bibinfo
  {journal} {Nat. Phys.}\ }\textbf {\bibinfo {volume} {17}},\ \bibinfo {pages}
  {332} (\bibinfo {year} {2021})}\BibitemShut {NoStop}%
\bibitem [{\citenamefont {Hornibrook}\ \emph {et~al.}(2015)\citenamefont
  {Hornibrook}, \citenamefont {Colless}, \citenamefont {Conway~Lamb},
  \citenamefont {Pauka}, \citenamefont {Lu}, \citenamefont {Gossard},
  \citenamefont {Watson}, \citenamefont {Gardner}, \citenamefont {Fallahi},
  \citenamefont {Manfra},\ and\ \citenamefont {Reilly}}]{Hornibrook2015}%
  \BibitemOpen
  \bibfield  {author} {\bibinfo {author} {\bibfnamefont {J.~M.}\ \bibnamefont
  {Hornibrook}}, \bibinfo {author} {\bibfnamefont {J.~I.}\ \bibnamefont
  {Colless}}, \bibinfo {author} {\bibfnamefont {I.~D.}\ \bibnamefont
  {Conway~Lamb}}, \bibinfo {author} {\bibfnamefont {S.~J.}\ \bibnamefont
  {Pauka}}, \bibinfo {author} {\bibfnamefont {H.}~\bibnamefont {Lu}}, \bibinfo
  {author} {\bibfnamefont {A.~C.}\ \bibnamefont {Gossard}}, \bibinfo {author}
  {\bibfnamefont {J.~D.}\ \bibnamefont {Watson}}, \bibinfo {author}
  {\bibfnamefont {G.~C.}\ \bibnamefont {Gardner}}, \bibinfo {author}
  {\bibfnamefont {S.}~\bibnamefont {Fallahi}}, \bibinfo {author} {\bibfnamefont
  {M.~J.}\ \bibnamefont {Manfra}},\ and\ \bibinfo {author} {\bibfnamefont
  {D.~J.}\ \bibnamefont {Reilly}},\ }\bibfield  {title} {\bibinfo {title}
  {Cryogenic control architecture for large-scale quantum computing},\ }\href
  {https://doi.org/10.1103/PhysRevApplied.3.024010} {\bibfield  {journal}
  {\bibinfo  {journal} {Phys. Rev. Applied}\ }\textbf {\bibinfo {volume} {3}},\
  \bibinfo {pages} {024010} (\bibinfo {year} {2015})}\BibitemShut {NoStop}%
\bibitem [{\citenamefont {Reilly}(2015)}]{Reilly2015}%
  \BibitemOpen
  \bibfield  {author} {\bibinfo {author} {\bibfnamefont {D.~J.}\ \bibnamefont
  {Reilly}},\ }\bibfield  {title} {\bibinfo {title} {{Engineering the
  quantum-classical interface of solid-state qubits}},\ }\href
  {https://doi.org/10.1038/npjqi.2015.11} {\bibfield  {journal} {\bibinfo
  {journal} {npj Quant. Inf.}\ }\textbf {\bibinfo {volume} {1}},\ \bibinfo
  {pages} {15011} (\bibinfo {year} {2015})}\BibitemShut {NoStop}%
\bibitem [{\citenamefont {Vandersypen}\ \emph {et~al.}(2017)\citenamefont
  {Vandersypen}, \citenamefont {Bluhm}, \citenamefont {Clarke}, \citenamefont
  {Dzurak}, \citenamefont {Ishihara}, \citenamefont {Morello}, \citenamefont
  {Reilly}, \citenamefont {Schreiber},\ and\ \citenamefont
  {Veldhorst}}]{Vandersypen2017}%
  \BibitemOpen
  \bibfield  {author} {\bibinfo {author} {\bibfnamefont {L.~M.~K.}\
  \bibnamefont {Vandersypen}}, \bibinfo {author} {\bibfnamefont
  {H.}~\bibnamefont {Bluhm}}, \bibinfo {author} {\bibfnamefont {J.~S.}\
  \bibnamefont {Clarke}}, \bibinfo {author} {\bibfnamefont {A.~S.}\
  \bibnamefont {Dzurak}}, \bibinfo {author} {\bibfnamefont {R.}~\bibnamefont
  {Ishihara}}, \bibinfo {author} {\bibfnamefont {A.}~\bibnamefont {Morello}},
  \bibinfo {author} {\bibfnamefont {D.~J.}\ \bibnamefont {Reilly}}, \bibinfo
  {author} {\bibfnamefont {L.~R.}\ \bibnamefont {Schreiber}},\ and\ \bibinfo
  {author} {\bibfnamefont {M.}~\bibnamefont {Veldhorst}},\ }\bibfield  {title}
  {\bibinfo {title} {{Interfacing spin qubits in quantum dots and donors—hot,
  dense, and coherent}},\ }\href
  {https://www.nature.com/articles/s41534-017-0038-y} {\bibfield  {journal}
  {\bibinfo  {journal} {npj Quant. Inf.}\ }\textbf {\bibinfo {volume} {3}},\
  \bibinfo {pages} {34} (\bibinfo {year} {2017})}\BibitemShut {NoStop}%
\bibitem [{\citenamefont {Bardin}\ \emph {et~al.}(2019)\citenamefont {Bardin},
  \citenamefont {White}, \citenamefont {Giustina}, \citenamefont {Satzinger},
  \citenamefont {Arya}, \citenamefont {Roushan}, \citenamefont {Chiaro},
  \citenamefont {Kelly}, \citenamefont {Chen}, \citenamefont {Burkett},
  \citenamefont {Chen}, \citenamefont {Jeffrey}, \citenamefont {Dunsworth},
  \citenamefont {Fowler}, \citenamefont {Foxen}, \citenamefont {Gidney},
  \citenamefont {Graff}, \citenamefont {Klimov}, \citenamefont {Mutus},
  \citenamefont {McEwen}, \citenamefont {Neeley}, \citenamefont {Neill},
  \citenamefont {Lucero}, \citenamefont {Quintana}, \citenamefont
  {Vainsencher}, \citenamefont {Neven}, \citenamefont {Martinis}, \citenamefont
  {Huang}, \citenamefont {Das}, \citenamefont {Sank}, \citenamefont {Naaman},
  \citenamefont {Megrant},\ and\ \citenamefont {Barends}}]{Bardin2019}%
  \BibitemOpen
  \bibfield  {author} {\bibinfo {author} {\bibfnamefont {J.~C.}\ \bibnamefont
  {Bardin}}, \bibinfo {author} {\bibfnamefont {T.}~\bibnamefont {White}},
  \bibinfo {author} {\bibfnamefont {M.}~\bibnamefont {Giustina}}, \bibinfo
  {author} {\bibfnamefont {K.~J.}\ \bibnamefont {Satzinger}}, \bibinfo {author}
  {\bibfnamefont {K.}~\bibnamefont {Arya}}, \bibinfo {author} {\bibfnamefont
  {P.}~\bibnamefont {Roushan}}, \bibinfo {author} {\bibfnamefont
  {B.}~\bibnamefont {Chiaro}}, \bibinfo {author} {\bibfnamefont
  {J.}~\bibnamefont {Kelly}}, \bibinfo {author} {\bibfnamefont
  {Z.}~\bibnamefont {Chen}}, \bibinfo {author} {\bibfnamefont {B.}~\bibnamefont
  {Burkett}}, \bibinfo {author} {\bibfnamefont {Y.}~\bibnamefont {Chen}},
  \bibinfo {author} {\bibfnamefont {E.}~\bibnamefont {Jeffrey}}, \bibinfo
  {author} {\bibfnamefont {A.}~\bibnamefont {Dunsworth}}, \bibinfo {author}
  {\bibfnamefont {A.}~\bibnamefont {Fowler}}, \bibinfo {author} {\bibfnamefont
  {B.}~\bibnamefont {Foxen}}, \bibinfo {author} {\bibfnamefont
  {C.}~\bibnamefont {Gidney}}, \bibinfo {author} {\bibfnamefont
  {R.}~\bibnamefont {Graff}}, \bibinfo {author} {\bibfnamefont
  {P.}~\bibnamefont {Klimov}}, \bibinfo {author} {\bibfnamefont
  {J.}~\bibnamefont {Mutus}}, \bibinfo {author} {\bibfnamefont {M.~J.}\
  \bibnamefont {McEwen}}, \bibinfo {author} {\bibfnamefont {M.}~\bibnamefont
  {Neeley}}, \bibinfo {author} {\bibfnamefont {C.~J.}\ \bibnamefont {Neill}},
  \bibinfo {author} {\bibfnamefont {E.}~\bibnamefont {Lucero}}, \bibinfo
  {author} {\bibfnamefont {C.}~\bibnamefont {Quintana}}, \bibinfo {author}
  {\bibfnamefont {A.}~\bibnamefont {Vainsencher}}, \bibinfo {author}
  {\bibfnamefont {H.}~\bibnamefont {Neven}}, \bibinfo {author} {\bibfnamefont
  {J.}~\bibnamefont {Martinis}}, \bibinfo {author} {\bibfnamefont
  {T.}~\bibnamefont {Huang}}, \bibinfo {author} {\bibfnamefont
  {S.}~\bibnamefont {Das}}, \bibinfo {author} {\bibfnamefont {D.~T.}\
  \bibnamefont {Sank}}, \bibinfo {author} {\bibfnamefont {O.}~\bibnamefont
  {Naaman}}, \bibinfo {author} {\bibfnamefont {A.~E.}\ \bibnamefont
  {Megrant}},\ and\ \bibinfo {author} {\bibfnamefont {R.}~\bibnamefont
  {Barends}},\ }\bibfield  {title} {\bibinfo {title} {{Design and
  Characterization of a 28-nm Bulk-CMOS Cryogenic Quantum Controller
  Dissipating Less Than 2 mW at 3 K}},\ }\href
  {https://doi.org/10.1109/jssc.2019.2937234} {\bibfield  {journal} {\bibinfo
  {journal} {IEEE J. Solid-State Circuits}\ }\textbf {\bibinfo {volume} {54}},\
  \bibinfo {pages} {3043} (\bibinfo {year} {2019})}\BibitemShut {NoStop}%
\bibitem [{\citenamefont {van Dijk}\ \emph {et~al.}(2019)\citenamefont {van
  Dijk}, \citenamefont {Kawakami}, \citenamefont {Schouten}, \citenamefont
  {Veldhorst}, \citenamefont {Vandersypen}, \citenamefont {Babaie},
  \citenamefont {Charbon},\ and\ \citenamefont {Sebastiano}}]{Dijk2019}%
  \BibitemOpen
  \bibfield  {author} {\bibinfo {author} {\bibfnamefont {J.}~\bibnamefont {van
  Dijk}}, \bibinfo {author} {\bibfnamefont {E.}~\bibnamefont {Kawakami}},
  \bibinfo {author} {\bibfnamefont {R.}~\bibnamefont {Schouten}}, \bibinfo
  {author} {\bibfnamefont {M.}~\bibnamefont {Veldhorst}}, \bibinfo {author}
  {\bibfnamefont {L.}~\bibnamefont {Vandersypen}}, \bibinfo {author}
  {\bibfnamefont {M.}~\bibnamefont {Babaie}}, \bibinfo {author} {\bibfnamefont
  {E.}~\bibnamefont {Charbon}},\ and\ \bibinfo {author} {\bibfnamefont
  {F.}~\bibnamefont {Sebastiano}},\ }\bibfield  {title} {\bibinfo {title}
  {Impact of classical control electronics on qubit fidelity},\ }\href
  {https://doi.org/10.1103/PhysRevApplied.12.044054} {\bibfield  {journal}
  {\bibinfo  {journal} {Phys. Rev. Applied}\ }\textbf {\bibinfo {volume}
  {12}},\ \bibinfo {pages} {044054} (\bibinfo {year} {2019})}\BibitemShut
  {NoStop}%
\bibitem [{\citenamefont {Petit}\ \emph {et~al.}(2020)\citenamefont {Petit},
  \citenamefont {Eenink}, \citenamefont {Russ}, \citenamefont {Lawrie},
  \citenamefont {Hendrickx}, \citenamefont {Philips}, \citenamefont {Clarke},
  \citenamefont {Vandersypen},\ and\ \citenamefont {Veldhorst}}]{Petit2020}%
  \BibitemOpen
  \bibfield  {author} {\bibinfo {author} {\bibfnamefont {L.}~\bibnamefont
  {Petit}}, \bibinfo {author} {\bibfnamefont {H.~G.~J.}\ \bibnamefont
  {Eenink}}, \bibinfo {author} {\bibfnamefont {M.}~\bibnamefont {Russ}},
  \bibinfo {author} {\bibfnamefont {W.~I.~L.}\ \bibnamefont {Lawrie}}, \bibinfo
  {author} {\bibfnamefont {N.~W.}\ \bibnamefont {Hendrickx}}, \bibinfo {author}
  {\bibfnamefont {S.~G.~J.}\ \bibnamefont {Philips}}, \bibinfo {author}
  {\bibfnamefont {J.~S.}\ \bibnamefont {Clarke}}, \bibinfo {author}
  {\bibfnamefont {L.~M.~K.}\ \bibnamefont {Vandersypen}},\ and\ \bibinfo
  {author} {\bibfnamefont {M.}~\bibnamefont {Veldhorst}},\ }\bibfield  {title}
  {\bibinfo {title} {{Universal quantum logic in hot silicon qubits}},\ }\href
  {https://doi.org/10.1038/s41586-020-2170-7} {\bibfield  {journal} {\bibinfo
  {journal} {Nature}\ }\textbf {\bibinfo {volume} {580}},\ \bibinfo {pages}
  {355} (\bibinfo {year} {2020})}\BibitemShut {NoStop}%
\bibitem [{\citenamefont {Pauka}\ \emph {et~al.}(2021)\citenamefont {Pauka},
  \citenamefont {Das}, \citenamefont {Kalra}, \citenamefont {Moini},
  \citenamefont {Yang}, \citenamefont {Trainer}, \citenamefont {Bousquet},
  \citenamefont {Cantaloube}, \citenamefont {Dick}, \citenamefont {Gardner},
  \citenamefont {Manfra},\ and\ \citenamefont {Reilly}}]{Pauka2021}%
  \BibitemOpen
  \bibfield  {author} {\bibinfo {author} {\bibfnamefont {S.~J.}\ \bibnamefont
  {Pauka}}, \bibinfo {author} {\bibfnamefont {K.}~\bibnamefont {Das}}, \bibinfo
  {author} {\bibfnamefont {R.}~\bibnamefont {Kalra}}, \bibinfo {author}
  {\bibfnamefont {A.}~\bibnamefont {Moini}}, \bibinfo {author} {\bibfnamefont
  {Y.}~\bibnamefont {Yang}}, \bibinfo {author} {\bibfnamefont {M.}~\bibnamefont
  {Trainer}}, \bibinfo {author} {\bibfnamefont {A.}~\bibnamefont {Bousquet}},
  \bibinfo {author} {\bibfnamefont {C.}~\bibnamefont {Cantaloube}}, \bibinfo
  {author} {\bibfnamefont {N.}~\bibnamefont {Dick}}, \bibinfo {author}
  {\bibfnamefont {G.~C.}\ \bibnamefont {Gardner}}, \bibinfo {author}
  {\bibfnamefont {M.~J.}\ \bibnamefont {Manfra}},\ and\ \bibinfo {author}
  {\bibfnamefont {D.~J.}\ \bibnamefont {Reilly}},\ }\bibfield  {title}
  {\bibinfo {title} {A cryogenic {CMOS} chip for generating control signals for
  multiple qubits},\ }\href {https://doi.org/10.1038/s41928-020-00528-y}
  {\bibfield  {journal} {\bibinfo  {journal} {Nat. Electron.}\ }\textbf
  {\bibinfo {volume} {4}},\ \bibinfo {pages} {64} (\bibinfo {year}
  {2021})}\BibitemShut {NoStop}%
\bibitem [{\citenamefont {Xue}\ \emph {et~al.}(2021)\citenamefont {Xue},
  \citenamefont {Patra}, \citenamefont {van Dijk}, \citenamefont {Samkharadze},
  \citenamefont {Subramanian}, \citenamefont {Corna}, \citenamefont {Wuetz},
  \citenamefont {Jeon}, \citenamefont {Sheikh}, \citenamefont
  {Juarez-Hernandez}, \citenamefont {Esparza}, \citenamefont {Rampurawala},
  \citenamefont {Carlton}, \citenamefont {Ravikumar}, \citenamefont {Nieva},
  \citenamefont {Kim}, \citenamefont {Lee}, \citenamefont {Sammak},
  \citenamefont {Scappucci}, \citenamefont {Veldhorst}, \citenamefont
  {Sebastiano}, \citenamefont {Babaie}, \citenamefont {Pellerano},
  \citenamefont {Charbon},\ and\ \citenamefont {Vandersypen}}]{Xue2021}%
  \BibitemOpen
  \bibfield  {author} {\bibinfo {author} {\bibfnamefont {X.}~\bibnamefont
  {Xue}}, \bibinfo {author} {\bibfnamefont {B.}~\bibnamefont {Patra}}, \bibinfo
  {author} {\bibfnamefont {J.~P.~G.}\ \bibnamefont {van Dijk}}, \bibinfo
  {author} {\bibfnamefont {N.}~\bibnamefont {Samkharadze}}, \bibinfo {author}
  {\bibfnamefont {S.}~\bibnamefont {Subramanian}}, \bibinfo {author}
  {\bibfnamefont {A.}~\bibnamefont {Corna}}, \bibinfo {author} {\bibfnamefont
  {B.~P.}\ \bibnamefont {Wuetz}}, \bibinfo {author} {\bibfnamefont
  {C.}~\bibnamefont {Jeon}}, \bibinfo {author} {\bibfnamefont {F.}~\bibnamefont
  {Sheikh}}, \bibinfo {author} {\bibfnamefont {E.}~\bibnamefont
  {Juarez-Hernandez}}, \bibinfo {author} {\bibfnamefont {B.~P.}\ \bibnamefont
  {Esparza}}, \bibinfo {author} {\bibfnamefont {H.}~\bibnamefont
  {Rampurawala}}, \bibinfo {author} {\bibfnamefont {B.}~\bibnamefont
  {Carlton}}, \bibinfo {author} {\bibfnamefont {S.}~\bibnamefont {Ravikumar}},
  \bibinfo {author} {\bibfnamefont {C.}~\bibnamefont {Nieva}}, \bibinfo
  {author} {\bibfnamefont {S.}~\bibnamefont {Kim}}, \bibinfo {author}
  {\bibfnamefont {H.-J.}\ \bibnamefont {Lee}}, \bibinfo {author} {\bibfnamefont
  {A.}~\bibnamefont {Sammak}}, \bibinfo {author} {\bibfnamefont
  {G.}~\bibnamefont {Scappucci}}, \bibinfo {author} {\bibfnamefont
  {M.}~\bibnamefont {Veldhorst}}, \bibinfo {author} {\bibfnamefont
  {F.}~\bibnamefont {Sebastiano}}, \bibinfo {author} {\bibfnamefont
  {M.}~\bibnamefont {Babaie}}, \bibinfo {author} {\bibfnamefont
  {S.}~\bibnamefont {Pellerano}}, \bibinfo {author} {\bibfnamefont
  {E.}~\bibnamefont {Charbon}},\ and\ \bibinfo {author} {\bibfnamefont
  {L.~M.~K.}\ \bibnamefont {Vandersypen}},\ }\bibfield  {title} {\bibinfo
  {title} {{CMOS}-based cryogenic control of silicon quantum circuits},\ }\href
  {https://doi.org/10.1038/s41586-021-03469-4} {\bibfield  {journal} {\bibinfo
  {journal} {Nature}\ }\textbf {\bibinfo {volume} {593}},\ \bibinfo {pages}
  {205} (\bibinfo {year} {2021})}\BibitemShut {NoStop}%
\bibitem [{\citenamefont {Chen}(2017)}]{Chen2017}%
  \BibitemOpen
  \bibfield  {author} {\bibinfo {author} {\bibfnamefont {Q.-M.}\ \bibnamefont
  {Chen}},\ }\emph {\bibinfo {title} {Pulse Width Modulation in Quantum
  Engineering}},\ \href@noop {} {Master's thesis},\ \bibinfo  {school}
  {Tsinghua University} (\bibinfo {year} {2017})\BibitemShut {NoStop}%
\bibitem [{\citenamefont {Holmes}\ and\ \citenamefont
  {Lipo}(2003)}]{Holmes2003}%
  \BibitemOpen
  \bibfield  {author} {\bibinfo {author} {\bibfnamefont {D.~G.}\ \bibnamefont
  {Holmes}}\ and\ \bibinfo {author} {\bibfnamefont {T.~A.}\ \bibnamefont
  {Lipo}},\ }\href@noop {} {\emph {\bibinfo {title} {Pulse width modulation for
  power converters: principles and practice}}},\ Vol.~\bibinfo {volume} {18}\
  (\bibinfo  {publisher} {John Wiley \& Sons},\ \bibinfo {year}
  {2003})\BibitemShut {NoStop}%
\bibitem [{sup()}]{supplementary}%
  \BibitemOpen
  \href@noop {} {\emph {\bibinfo {title} {\rm See Supplementary Materials for
  additional details}}}\BibitemShut {NoStop}%
\bibitem [{\citenamefont {Moler}\ and\ \citenamefont
  {Van~Loan}(2003)}]{Moler2003}%
  \BibitemOpen
  \bibfield  {author} {\bibinfo {author} {\bibfnamefont {C.}~\bibnamefont
  {Moler}}\ and\ \bibinfo {author} {\bibfnamefont {C.}~\bibnamefont
  {Van~Loan}},\ }\bibfield  {title} {\bibinfo {title} {Nineteen dubious ways to
  compute the exponential of a matrix, twenty-five years later},\ }\href
  {https://doi.org/10.1137/S00361445024180} {\bibfield  {journal} {\bibinfo
  {journal} {SIAM Rev.}\ }\textbf {\bibinfo {volume} {45}},\ \bibinfo {pages}
  {3} (\bibinfo {year} {2003})}\BibitemShut {NoStop}%
\bibitem [{\citenamefont {Fleck}\ \emph {et~al.}(1976)\citenamefont {Fleck},
  \citenamefont {Morris},\ and\ \citenamefont {Feit}}]{Fleck1976}%
  \BibitemOpen
  \bibfield  {author} {\bibinfo {author} {\bibfnamefont {J.~A.}\ \bibnamefont
  {Fleck}}, \bibinfo {author} {\bibfnamefont {J.~R.}\ \bibnamefont {Morris}},\
  and\ \bibinfo {author} {\bibfnamefont {M.~D.}\ \bibnamefont {Feit}},\
  }\bibfield  {title} {\bibinfo {title} {Time-dependent propagation of high
  energy laser beams through the atmosphere},\ }\href
  {https://doi.org/10.1007/BF00896333} {\bibfield  {journal} {\bibinfo
  {journal} {Appl. Phys.}\ }\textbf {\bibinfo {volume} {10}},\ \bibinfo {pages}
  {129} (\bibinfo {year} {1976})}\BibitemShut {NoStop}%
\bibitem [{\citenamefont {Feit}\ \emph {et~al.}(1982)\citenamefont {Feit},
  \citenamefont {Fleck},\ and\ \citenamefont {Steiger}}]{Feit1982}%
  \BibitemOpen
  \bibfield  {author} {\bibinfo {author} {\bibfnamefont {M.}~\bibnamefont
  {Feit}}, \bibinfo {author} {\bibfnamefont {J.}~\bibnamefont {Fleck}},\ and\
  \bibinfo {author} {\bibfnamefont {A.}~\bibnamefont {Steiger}},\ }\bibfield
  {title} {\bibinfo {title} {Solution of the schrödinger equation by a
  spectral method},\ }\href
  {https://doi.org/https://doi.org/10.1016/0021-9991(82)90091-2} {\bibfield
  {journal} {\bibinfo  {journal} {J. Comput. Phys.}\ }\textbf {\bibinfo
  {volume} {47}},\ \bibinfo {pages} {412 } (\bibinfo {year}
  {1982})}\BibitemShut {NoStop}%
\bibitem [{\citenamefont {Feit}\ and\ \citenamefont {Fleck}(1983)}]{Feit1983}%
  \BibitemOpen
  \bibfield  {author} {\bibinfo {author} {\bibfnamefont {M.~D.}\ \bibnamefont
  {Feit}}\ and\ \bibinfo {author} {\bibfnamefont {J.~A.}\ \bibnamefont
  {Fleck}},\ }\bibfield  {title} {\bibinfo {title} {Solution of the
  schrödinger equation by a spectral method ii: Vibrational energy levels of
  triatomic molecules},\ }\href {https://doi.org/10.1063/1.444501} {\bibfield
  {journal} {\bibinfo  {journal} {J. Chem. Phys.}\ }\textbf {\bibinfo {volume}
  {78}},\ \bibinfo {pages} {301} (\bibinfo {year} {1983})}\BibitemShut
  {NoStop}%
\bibitem [{\citenamefont {Ding}\ and\ \citenamefont {Wu}(2019)}]{Ding2019}%
  \BibitemOpen
  \bibfield  {author} {\bibinfo {author} {\bibfnamefont {H.-J.}\ \bibnamefont
  {Ding}}\ and\ \bibinfo {author} {\bibfnamefont {R.-B.}\ \bibnamefont {Wu}},\
  }\bibfield  {title} {\bibinfo {title} {Robust quantum control against clock
  noises in multiqubit systems},\ }\href
  {https://doi.org/10.1103/PhysRevA.100.022302} {\bibfield  {journal} {\bibinfo
   {journal} {Phys. Rev. A}\ }\textbf {\bibinfo {volume} {100}},\ \bibinfo
  {pages} {022302} (\bibinfo {year} {2019})}\BibitemShut {NoStop}%
\bibitem [{Note1()}]{Note1}%
  \BibitemOpen
  \bibinfo {note} {This value is estimated by noticing that $0.9999^{1000}
  \approx 0.9$.}\BibitemShut {Stop}%
\bibitem [{\citenamefont {Viola}\ \emph
  {et~al.}(1999{\natexlab{a}})\citenamefont {Viola}, \citenamefont {Knill},\
  and\ \citenamefont {Lloyd}}]{Viola1999}%
  \BibitemOpen
  \bibfield  {author} {\bibinfo {author} {\bibfnamefont {L.}~\bibnamefont
  {Viola}}, \bibinfo {author} {\bibfnamefont {E.}~\bibnamefont {Knill}},\ and\
  \bibinfo {author} {\bibfnamefont {S.}~\bibnamefont {Lloyd}},\ }\bibfield
  {title} {\bibinfo {title} {Dynamical decoupling of open quantum systems},\
  }\href {https://doi.org/10.1103/PhysRevLett.82.2417} {\bibfield  {journal}
  {\bibinfo  {journal} {Phys. Rev. Lett.}\ }\textbf {\bibinfo {volume} {82}},\
  \bibinfo {pages} {2417} (\bibinfo {year} {1999}{\natexlab{a}})}\BibitemShut
  {NoStop}%
\bibitem [{\citenamefont {Viola}\ \emph
  {et~al.}(1999{\natexlab{b}})\citenamefont {Viola}, \citenamefont {Lloyd},\
  and\ \citenamefont {Knill}}]{Viola1999a}%
  \BibitemOpen
  \bibfield  {author} {\bibinfo {author} {\bibfnamefont {L.}~\bibnamefont
  {Viola}}, \bibinfo {author} {\bibfnamefont {S.}~\bibnamefont {Lloyd}},\ and\
  \bibinfo {author} {\bibfnamefont {E.}~\bibnamefont {Knill}},\ }\bibfield
  {title} {\bibinfo {title} {Universal control of decoupled quantum systems},\
  }\href {https://doi.org/10.1103/PhysRevLett.83.4888} {\bibfield  {journal}
  {\bibinfo  {journal} {Phys. Rev. Lett.}\ }\textbf {\bibinfo {volume} {83}},\
  \bibinfo {pages} {4888} (\bibinfo {year} {1999}{\natexlab{b}})}\BibitemShut
  {NoStop}%
\bibitem [{\citenamefont {Lloyd}\ and\ \citenamefont
  {Viola}(2001)}]{Lloyd2001}%
  \BibitemOpen
  \bibfield  {author} {\bibinfo {author} {\bibfnamefont {S.}~\bibnamefont
  {Lloyd}}\ and\ \bibinfo {author} {\bibfnamefont {L.}~\bibnamefont {Viola}},\
  }\bibfield  {title} {\bibinfo {title} {Engineering quantum dynamics},\ }\href
  {https://doi.org/10.1103/PhysRevA.65.010101} {\bibfield  {journal} {\bibinfo
  {journal} {Phys. Rev. A}\ }\textbf {\bibinfo {volume} {65}},\ \bibinfo
  {pages} {010101} (\bibinfo {year} {2001})}\BibitemShut {NoStop}%
\bibitem [{\citenamefont {Trotter}(1959)}]{Trotter1959}%
  \BibitemOpen
  \bibfield  {author} {\bibinfo {author} {\bibfnamefont {H.~F.}\ \bibnamefont
  {Trotter}},\ }\bibfield  {title} {\bibinfo {title} {On the product of
  semi-groups of operators},\ }\href
  {https://doi.org/10.1090/S0002-9939-1959-0108732-6} {\bibfield  {journal}
  {\bibinfo  {journal} {Proc. Amer. Math. Soc.}\ }\textbf {\bibinfo {volume}
  {10}},\ \bibinfo {pages} {545} (\bibinfo {year} {1959})}\BibitemShut
  {NoStop}%
\bibitem [{\citenamefont {Suzuki}(1985)}]{Suzuki1985}%
  \BibitemOpen
  \bibfield  {author} {\bibinfo {author} {\bibfnamefont {M.}~\bibnamefont
  {Suzuki}},\ }\bibfield  {title} {\bibinfo {title} {Decomposition formulas of
  exponential operators and lie exponentials with some applications to quantum
  mechanics and statistical physics},\ }\href
  {https://doi.org/10.1063/1.526596} {\bibfield  {journal} {\bibinfo  {journal}
  {J. Math. Phys.}\ }\textbf {\bibinfo {volume} {26}},\ \bibinfo {pages} {601}
  (\bibinfo {year} {1985})}\BibitemShut {NoStop}%
\bibitem [{\citenamefont {Bandrauk}\ and\ \citenamefont
  {Shen}(1992)}]{Bandrauk1992}%
  \BibitemOpen
  \bibfield  {author} {\bibinfo {author} {\bibfnamefont {A.~D.}\ \bibnamefont
  {Bandrauk}}\ and\ \bibinfo {author} {\bibfnamefont {H.}~\bibnamefont
  {Shen}},\ }\bibfield  {title} {\bibinfo {title} {Higher order exponential
  split operator method for solving time-dependent schrödinger equations},\
  }\href {https://doi.org/10.1139/v92-078} {\bibfield  {journal} {\bibinfo
  {journal} {Can. J. Chem.}\ }\textbf {\bibinfo {volume} {70}},\ \bibinfo
  {pages} {555} (\bibinfo {year} {1992})}\BibitemShut {NoStop}%
\bibitem [{\citenamefont {Bandrauk}\ and\ \citenamefont
  {Shen}(1993)}]{Bandrauk1993}%
  \BibitemOpen
  \bibfield  {author} {\bibinfo {author} {\bibfnamefont {A.~D.}\ \bibnamefont
  {Bandrauk}}\ and\ \bibinfo {author} {\bibfnamefont {H.}~\bibnamefont
  {Shen}},\ }\bibfield  {title} {\bibinfo {title} {Exponential split operator
  methods for solving coupled time‐dependent schrödinger equations},\ }\href
  {https://doi.org/10.1063/1.465362} {\bibfield  {journal} {\bibinfo  {journal}
  {J. Chem. Phys.}\ }\textbf {\bibinfo {volume} {99}},\ \bibinfo {pages} {1185}
  (\bibinfo {year} {1993})}\BibitemShut {NoStop}%
\bibitem [{\citenamefont {Zanardi}\ and\ \citenamefont
  {Lidar}(2004)}]{Zanardi2004}%
  \BibitemOpen
  \bibfield  {author} {\bibinfo {author} {\bibfnamefont {P.}~\bibnamefont
  {Zanardi}}\ and\ \bibinfo {author} {\bibfnamefont {D.~A.}\ \bibnamefont
  {Lidar}},\ }\bibfield  {title} {\bibinfo {title} {Purity and state fidelity
  of quantum channels},\ }\href {https://doi.org/10.1103/PhysRevA.70.012315}
  {\bibfield  {journal} {\bibinfo  {journal} {Phys. Rev. A}\ }\textbf {\bibinfo
  {volume} {70}},\ \bibinfo {pages} {012315} (\bibinfo {year}
  {2004})}\BibitemShut {NoStop}%
\bibitem [{\citenamefont {Pedersen}\ \emph {et~al.}(2007)\citenamefont
  {Pedersen}, \citenamefont {M{\o}ller},\ and\ \citenamefont
  {M{\o}lmer}}]{Pedersen2007}%
  \BibitemOpen
  \bibfield  {author} {\bibinfo {author} {\bibfnamefont {L.~H.}\ \bibnamefont
  {Pedersen}}, \bibinfo {author} {\bibfnamefont {N.~M.}\ \bibnamefont
  {M{\o}ller}},\ and\ \bibinfo {author} {\bibfnamefont {K.}~\bibnamefont
  {M{\o}lmer}},\ }\bibfield  {title} {\bibinfo {title} {Fidelity of quantum
  operations},\ }\href
  {https://doi.org/https://doi.org/10.1016/j.physleta.2007.02.069} {\bibfield
  {journal} {\bibinfo  {journal} {Phys. Lett. A}\ }\textbf {\bibinfo {volume}
  {367}},\ \bibinfo {pages} {47} (\bibinfo {year} {2007})}\BibitemShut
  {NoStop}%
\bibitem [{Note2()}]{Note2}%
  \BibitemOpen
  \bibinfo {note} {We note that all the converted waveforms for AWG
  implementation can be further optimized to $0.9999$ fidelity with
  conventional QOC algorithms.}\BibitemShut {Stop}%
\bibitem [{\citenamefont {{Likharev}}\ and\ \citenamefont
  {{Semenov}}(1991)}]{Likharev1991}%
  \BibitemOpen
  \bibfield  {author} {\bibinfo {author} {\bibfnamefont {K.~K.}\ \bibnamefont
  {{Likharev}}}\ and\ \bibinfo {author} {\bibfnamefont {V.~K.}\ \bibnamefont
  {{Semenov}}},\ }\bibfield  {title} {\bibinfo {title} {{RSFQ} logic/memory
  family: a new josephson-junction technology for sub-terahertz-clock-frequency
  digital systems},\ }\href {https://doi.org/10.1109/77.80745} {\bibfield
  {journal} {\bibinfo  {journal} {IEEE Trans. Appl. Supercond.}\ }\textbf
  {\bibinfo {volume} {1}},\ \bibinfo {pages} {3} (\bibinfo {year}
  {1991})}\BibitemShut {NoStop}%
\bibitem [{\citenamefont {{Zhou}}\ \emph {et~al.}(2001)\citenamefont {{Zhou}},
  \citenamefont {{Habif}}, \citenamefont {{Herr}}, \citenamefont {{Feldman}},\
  and\ \citenamefont {{Bocko}}}]{Zhou2001}%
  \BibitemOpen
  \bibfield  {author} {\bibinfo {author} {\bibfnamefont {X.}~\bibnamefont
  {{Zhou}}}, \bibinfo {author} {\bibfnamefont {J.~L.}\ \bibnamefont {{Habif}}},
  \bibinfo {author} {\bibfnamefont {A.~M.}\ \bibnamefont {{Herr}}}, \bibinfo
  {author} {\bibfnamefont {M.~J.}\ \bibnamefont {{Feldman}}},\ and\ \bibinfo
  {author} {\bibfnamefont {M.~F.}\ \bibnamefont {{Bocko}}},\ }\bibfield
  {title} {\bibinfo {title} {A tipping pulse scheme for a rf-{SQUID} qubit},\
  }\href {https://doi.org/10.1109/77.919522} {\bibfield  {journal} {\bibinfo
  {journal} {IEEE Trans. Appl. Supercond.}\ }\textbf {\bibinfo {volume} {11}},\
  \bibinfo {pages} {1018} (\bibinfo {year} {2001})}\BibitemShut {NoStop}%
\bibitem [{\citenamefont {{Crankshaw}}\ \emph {et~al.}(2003)\citenamefont
  {{Crankshaw}}, \citenamefont {{Habif}}, , \citenamefont {{Orlando}},
  \citenamefont {{Feldman}},\ and\ \citenamefont {{Bocko}}}]{Crankshaw2003}%
  \BibitemOpen
  \bibfield  {author} {\bibinfo {author} {\bibfnamefont {D.~S.}\ \bibnamefont
  {{Crankshaw}}}, \bibinfo {author} {\bibfnamefont {J.~L.}\ \bibnamefont
  {{Habif}}}, , \bibinfo {author} {\bibfnamefont {T.~P.}\ \bibnamefont
  {{Orlando}}}, \bibinfo {author} {\bibfnamefont {M.~J.}\ \bibnamefont
  {{Feldman}}},\ and\ \bibinfo {author} {\bibfnamefont {M.~F.}\ \bibnamefont
  {{Bocko}}},\ }\bibfield  {title} {\bibinfo {title} {An {RSFQ} variable duty
  cycle oscillator for driving a superconductive qubit},\ }\href
  {https://doi.org/10.1109/TASC.2003.814115} {\bibfield  {journal} {\bibinfo
  {journal} {IEEE Trans. Appl. Supercond.}\ }\textbf {\bibinfo {volume} {13}},\
  \bibinfo {pages} {966} (\bibinfo {year} {2003})}\BibitemShut {NoStop}%
\bibitem [{\citenamefont {{Semenov}}\ and\ \citenamefont
  {{Averin}}(2003)}]{Semenov2003}%
  \BibitemOpen
  \bibfield  {author} {\bibinfo {author} {\bibfnamefont {V.~K.}\ \bibnamefont
  {{Semenov}}}\ and\ \bibinfo {author} {\bibfnamefont {D.~V.}\ \bibnamefont
  {{Averin}}},\ }\bibfield  {title} {\bibinfo {title} {{SFQ} control circuits
  for josephson junction qubits},\ }\href
  {https://doi.org/10.1109/TASC.2003.814114} {\bibfield  {journal} {\bibinfo
  {journal} {IEEE Trans. Appl. Supercond.}\ }\textbf {\bibinfo {volume} {13}},\
  \bibinfo {pages} {960} (\bibinfo {year} {2003})}\BibitemShut {NoStop}%
\bibitem [{\citenamefont {McDermott}\ and\ \citenamefont
  {Vavilov}(2014)}]{McDermott2014}%
  \BibitemOpen
  \bibfield  {author} {\bibinfo {author} {\bibfnamefont {R.}~\bibnamefont
  {McDermott}}\ and\ \bibinfo {author} {\bibfnamefont {M.~G.}\ \bibnamefont
  {Vavilov}},\ }\bibfield  {title} {\bibinfo {title} {Accurate qubit control
  with single flux quantum pulses},\ }\href
  {https://doi.org/10.1103/PhysRevApplied.2.014007} {\bibfield  {journal}
  {\bibinfo  {journal} {Phys. Rev. Applied}\ }\textbf {\bibinfo {volume} {2}},\
  \bibinfo {pages} {014007} (\bibinfo {year} {2014})}\BibitemShut {NoStop}%
\bibitem [{\citenamefont {Liebermann}\ and\ \citenamefont
  {Wilhelm}(2016)}]{Liebermann2016}%
  \BibitemOpen
  \bibfield  {author} {\bibinfo {author} {\bibfnamefont {P.~J.}\ \bibnamefont
  {Liebermann}}\ and\ \bibinfo {author} {\bibfnamefont {F.~K.}\ \bibnamefont
  {Wilhelm}},\ }\bibfield  {title} {\bibinfo {title} {Optimal qubit control
  using single-flux quantum pulses},\ }\href
  {https://doi.org/10.1103/PhysRevApplied.6.024022} {\bibfield  {journal}
  {\bibinfo  {journal} {Phys. Rev. Applied}\ }\textbf {\bibinfo {volume} {6}},\
  \bibinfo {pages} {024022} (\bibinfo {year} {2016})}\BibitemShut {NoStop}%
\bibitem [{\citenamefont {McDermott}\ \emph {et~al.}(2018)\citenamefont
  {McDermott}, \citenamefont {Vavilov}, \citenamefont {Plourde}, \citenamefont
  {Wilhelm}, \citenamefont {Liebermann}, \citenamefont {Mukhanov},\ and\
  \citenamefont {Ohki}}]{McDermott2018}%
  \BibitemOpen
  \bibfield  {author} {\bibinfo {author} {\bibfnamefont {R.}~\bibnamefont
  {McDermott}}, \bibinfo {author} {\bibfnamefont {M.~G.}\ \bibnamefont
  {Vavilov}}, \bibinfo {author} {\bibfnamefont {B.~L.~T.}\ \bibnamefont
  {Plourde}}, \bibinfo {author} {\bibfnamefont {F.~K.}\ \bibnamefont
  {Wilhelm}}, \bibinfo {author} {\bibfnamefont {P.~J.}\ \bibnamefont
  {Liebermann}}, \bibinfo {author} {\bibfnamefont {O.~A.}\ \bibnamefont
  {Mukhanov}},\ and\ \bibinfo {author} {\bibfnamefont {T.~A.}\ \bibnamefont
  {Ohki}},\ }\bibfield  {title} {\bibinfo {title} {{Quantum–classical
  interface based on single flux quantum digital logic}},\ }\href
  {https://doi.org/10.1088/2058-9565/aaa3a0} {\bibfield  {journal} {\bibinfo
  {journal} {Quantum Sci. Technol.}\ }\textbf {\bibinfo {volume} {3}},\
  \bibinfo {pages} {024004} (\bibinfo {year} {2018})}\BibitemShut {NoStop}%
\bibitem [{\citenamefont {Leonard}\ \emph {et~al.}(2019)\citenamefont
  {Leonard}, \citenamefont {Beck}, \citenamefont {Nelson}, \citenamefont
  {Christensen}, \citenamefont {Thorbeck}, \citenamefont {Howington},
  \citenamefont {Opremcak}, \citenamefont {Pechenezhskiy}, \citenamefont
  {Dodge}, \citenamefont {Dupuis}, \citenamefont {Hutchings}, \citenamefont
  {Ku}, \citenamefont {Schlenker}, \citenamefont {Suttle}, \citenamefont
  {Wilen}, \citenamefont {Zhu}, \citenamefont {Vavilov}, \citenamefont
  {Plourde},\ and\ \citenamefont {McDermott}}]{Leonard2019}%
  \BibitemOpen
  \bibfield  {author} {\bibinfo {author} {\bibfnamefont {E.}~\bibnamefont
  {Leonard}}, \bibinfo {author} {\bibfnamefont {M.~A.}\ \bibnamefont {Beck}},
  \bibinfo {author} {\bibfnamefont {J.}~\bibnamefont {Nelson}}, \bibinfo
  {author} {\bibfnamefont {B.}~\bibnamefont {Christensen}}, \bibinfo {author}
  {\bibfnamefont {T.}~\bibnamefont {Thorbeck}}, \bibinfo {author}
  {\bibfnamefont {C.}~\bibnamefont {Howington}}, \bibinfo {author}
  {\bibfnamefont {A.}~\bibnamefont {Opremcak}}, \bibinfo {author}
  {\bibfnamefont {I.}~\bibnamefont {Pechenezhskiy}}, \bibinfo {author}
  {\bibfnamefont {K.}~\bibnamefont {Dodge}}, \bibinfo {author} {\bibfnamefont
  {N.}~\bibnamefont {Dupuis}}, \bibinfo {author} {\bibfnamefont
  {M.}~\bibnamefont {Hutchings}}, \bibinfo {author} {\bibfnamefont
  {J.}~\bibnamefont {Ku}}, \bibinfo {author} {\bibfnamefont {F.}~\bibnamefont
  {Schlenker}}, \bibinfo {author} {\bibfnamefont {J.}~\bibnamefont {Suttle}},
  \bibinfo {author} {\bibfnamefont {C.}~\bibnamefont {Wilen}}, \bibinfo
  {author} {\bibfnamefont {S.}~\bibnamefont {Zhu}}, \bibinfo {author}
  {\bibfnamefont {M.}~\bibnamefont {Vavilov}}, \bibinfo {author} {\bibfnamefont
  {B.}~\bibnamefont {Plourde}},\ and\ \bibinfo {author} {\bibfnamefont
  {R.}~\bibnamefont {McDermott}},\ }\bibfield  {title} {\bibinfo {title}
  {Digital coherent control of a superconducting qubit},\ }\href
  {https://doi.org/10.1103/PhysRevApplied.11.014009} {\bibfield  {journal}
  {\bibinfo  {journal} {Phys. Rev. Applied}\ }\textbf {\bibinfo {volume}
  {11}},\ \bibinfo {pages} {014009} (\bibinfo {year} {2019})}\BibitemShut
  {NoStop}%
\bibitem [{\citenamefont {Li}\ \emph {et~al.}(2019)\citenamefont {Li},
  \citenamefont {McDermott},\ and\ \citenamefont {Vavilov}}]{Li2019}%
  \BibitemOpen
  \bibfield  {author} {\bibinfo {author} {\bibfnamefont {K.}~\bibnamefont
  {Li}}, \bibinfo {author} {\bibfnamefont {R.}~\bibnamefont {McDermott}},\ and\
  \bibinfo {author} {\bibfnamefont {M.~G.}\ \bibnamefont {Vavilov}},\
  }\bibfield  {title} {\bibinfo {title} {Hardware-efficient qubit control with
  single-flux-quantum pulse sequences},\ }\href
  {https://doi.org/10.1103/PhysRevApplied.12.014044} {\bibfield  {journal}
  {\bibinfo  {journal} {Phys. Rev. Applied}\ }\textbf {\bibinfo {volume}
  {12}},\ \bibinfo {pages} {014044} (\bibinfo {year} {2019})}\BibitemShut
  {NoStop}%
\bibitem [{\citenamefont {Khaneja}\ \emph {et~al.}(2001)\citenamefont
  {Khaneja}, \citenamefont {Brockett},\ and\ \citenamefont
  {Glaser}}]{Khaneja2001}%
  \BibitemOpen
  \bibfield  {author} {\bibinfo {author} {\bibfnamefont {N.}~\bibnamefont
  {Khaneja}}, \bibinfo {author} {\bibfnamefont {R.}~\bibnamefont {Brockett}},\
  and\ \bibinfo {author} {\bibfnamefont {S.~J.}\ \bibnamefont {Glaser}},\
  }\bibfield  {title} {\bibinfo {title} {Time optimal control in spin
  systems},\ }\href {https://doi.org/10.1103/PhysRevA.63.032308} {\bibfield
  {journal} {\bibinfo  {journal} {Phys. Rev. A}\ }\textbf {\bibinfo {volume}
  {63}},\ \bibinfo {pages} {032308} (\bibinfo {year} {2001})}\BibitemShut
  {NoStop}%
\bibitem [{\citenamefont {Chen}\ \emph {et~al.}(2015)\citenamefont {Chen},
  \citenamefont {Wu}, \citenamefont {Zhang},\ and\ \citenamefont
  {Rabitz}}]{Chen2015}%
  \BibitemOpen
  \bibfield  {author} {\bibinfo {author} {\bibfnamefont {Q.-M.}\ \bibnamefont
  {Chen}}, \bibinfo {author} {\bibfnamefont {R.-B.}\ \bibnamefont {Wu}},
  \bibinfo {author} {\bibfnamefont {T.-M.}\ \bibnamefont {Zhang}},\ and\
  \bibinfo {author} {\bibfnamefont {H.}~\bibnamefont {Rabitz}},\ }\bibfield
  {title} {\bibinfo {title} {Near-time-optimal control for quantum systems},\
  }\href {https://doi.org/10.1103/PhysRevA.92.063415} {\bibfield  {journal}
  {\bibinfo  {journal} {Phys. Rev. A}\ }\textbf {\bibinfo {volume} {92}},\
  \bibinfo {pages} {063415} (\bibinfo {year} {2015})}\BibitemShut {NoStop}%
\bibitem [{\citenamefont {Chen}\ \emph {et~al.}(2020)\citenamefont {Chen},
  \citenamefont {Yang}, \citenamefont {Arenz}, \citenamefont {Wu},
  \citenamefont {Peng}, \citenamefont {Pelczer},\ and\ \citenamefont
  {Rabitz}}]{Chen2020}%
  \BibitemOpen
  \bibfield  {author} {\bibinfo {author} {\bibfnamefont {Q.-M.}\ \bibnamefont
  {Chen}}, \bibinfo {author} {\bibfnamefont {X.}~\bibnamefont {Yang}}, \bibinfo
  {author} {\bibfnamefont {C.}~\bibnamefont {Arenz}}, \bibinfo {author}
  {\bibfnamefont {R.-B.}\ \bibnamefont {Wu}}, \bibinfo {author} {\bibfnamefont
  {X.}~\bibnamefont {Peng}}, \bibinfo {author} {\bibfnamefont {I.}~\bibnamefont
  {Pelczer}},\ and\ \bibinfo {author} {\bibfnamefont {H.}~\bibnamefont
  {Rabitz}},\ }\bibfield  {title} {\bibinfo {title} {Combining the synergistic
  control capabilities of modeling and experiments: Illustration of finding a
  minimum-time quantum objective},\ }\href
  {https://doi.org/10.1103/PhysRevA.101.032313} {\bibfield  {journal} {\bibinfo
   {journal} {Phys. Rev. A}\ }\textbf {\bibinfo {volume} {101}},\ \bibinfo
  {pages} {032313} (\bibinfo {year} {2020})}\BibitemShut {NoStop}%
\end{thebibliography}%


\begin{thebibliography}{5}%
\makeatletter
\providecommand \@ifxundefined [1]{%
 \@ifx{#1\undefined}
}%
\providecommand \@ifnum [1]{%
 \ifnum #1\expandafter \@firstoftwo
 \else \expandafter \@secondoftwo
 \fi
}%
\providecommand \@ifx [1]{%
 \ifx #1\expandafter \@firstoftwo
 \else \expandafter \@secondoftwo
 \fi
}%
\providecommand \natexlab [1]{#1}%
\providecommand \enquote  [1]{``#1''}%
\providecommand \bibnamefont  [1]{#1}%
\providecommand \bibfnamefont [1]{#1}%
\providecommand \citenamefont [1]{#1}%
\providecommand \href@noop [0]{\@secondoftwo}%
\providecommand \href [0]{\begingroup \@sanitize@url \@href}%
\providecommand \@href[1]{\@@startlink{#1}\@@href}%
\providecommand \@@href[1]{\endgroup#1\@@endlink}%
\providecommand \@sanitize@url [0]{\catcode `\\12\catcode `\$12\catcode
  `\&12\catcode `\#12\catcode `\^12\catcode `\_12\catcode `\%12\relax}%
\providecommand \@@startlink[1]{}%
\providecommand \@@endlink[0]{}%
\providecommand \url  [0]{\begingroup\@sanitize@url \@url }%
\providecommand \@url [1]{\endgroup\@href {#1}{\urlprefix }}%
\providecommand \urlprefix  [0]{URL }%
\providecommand \Eprint [0]{\href }%
\providecommand \doibase [0]{https://doi.org/}%
\providecommand \selectlanguage [0]{\@gobble}%
\providecommand \bibinfo  [0]{\@secondoftwo}%
\providecommand \bibfield  [0]{\@secondoftwo}%
\providecommand \translation [1]{[#1]}%
\providecommand \BibitemOpen [0]{}%
\providecommand \bibitemStop [0]{}%
\providecommand \bibitemNoStop [0]{.\EOS\space}%
\providecommand \EOS [0]{\spacefactor3000\relax}%
\providecommand \BibitemShut  [1]{\csname bibitem#1\endcsname}%
\let\auto@bib@innerbib\@empty
\bibitem [{\citenamefont {Moler}\ and\ \citenamefont
  {Van~Loan}(2003)}]{Moler2003}%
  \BibitemOpen
  \bibfield  {author} {\bibinfo {author} {\bibfnamefont {C.}~\bibnamefont
  {Moler}}\ and\ \bibinfo {author} {\bibfnamefont {C.}~\bibnamefont
  {Van~Loan}},\ }\bibfield  {title} {\bibinfo {title} {Nineteen dubious ways to
  compute the exponential of a matrix, twenty-five years later},\ }\href
  {https://doi.org/10.1137/S00361445024180} {\bibfield  {journal} {\bibinfo
  {journal} {SIAM Rev.}\ }\textbf {\bibinfo {volume} {45}},\ \bibinfo {pages}
  {3} (\bibinfo {year} {2003})}\BibitemShut {NoStop}%
\bibitem [{\citenamefont {Holmes}\ and\ \citenamefont
  {Lipo}(2003)}]{Holmes2003}%
  \BibitemOpen
  \bibfield  {author} {\bibinfo {author} {\bibfnamefont {D.~G.}\ \bibnamefont
  {Holmes}}\ and\ \bibinfo {author} {\bibfnamefont {T.~A.}\ \bibnamefont
  {Lipo}},\ }\href@noop {} {\emph {\bibinfo {title} {Pulse width modulation for
  power converters: principles and practice}}},\ Vol.~\bibinfo {volume} {18}\
  (\bibinfo  {publisher} {John Wiley \& Sons},\ \bibinfo {year}
  {2003})\BibitemShut {NoStop}%
\bibitem [{\citenamefont {Bandrauk}\ and\ \citenamefont
  {Shen}(1991)}]{Bandrauk1991}%
  \BibitemOpen
  \bibfield  {author} {\bibinfo {author} {\bibfnamefont {A.~D.}\ \bibnamefont
  {Bandrauk}}\ and\ \bibinfo {author} {\bibfnamefont {H.}~\bibnamefont
  {Shen}},\ }\bibfield  {title} {\bibinfo {title} {Improved exponential split
  operator method for solving the time-dependent schrödinger equation},\
  }\href {https://doi.org/https://doi.org/10.1016/0009-2614(91)90232-X}
  {\bibfield  {journal} {\bibinfo  {journal} {Chem. Phys. Lett.}\ }\textbf
  {\bibinfo {volume} {176}},\ \bibinfo {pages} {428 } (\bibinfo {year}
  {1991})}\BibitemShut {NoStop}%
\bibitem [{\citenamefont {Bandrauk}\ and\ \citenamefont
  {Shen}(1992)}]{Bandrauk1992}%
  \BibitemOpen
  \bibfield  {author} {\bibinfo {author} {\bibfnamefont {A.~D.}\ \bibnamefont
  {Bandrauk}}\ and\ \bibinfo {author} {\bibfnamefont {H.}~\bibnamefont
  {Shen}},\ }\bibfield  {title} {\bibinfo {title} {Higher order exponential
  split operator method for solving time-dependent schrödinger equations},\
  }\href {https://doi.org/10.1139/v92-078} {\bibfield  {journal} {\bibinfo
  {journal} {Can. J. Chem.}\ }\textbf {\bibinfo {volume} {70}},\ \bibinfo
  {pages} {555} (\bibinfo {year} {1992})}\BibitemShut {NoStop}%
\bibitem [{\citenamefont {Bandrauk}\ and\ \citenamefont
  {Shen}(1993)}]{Bandrauk1993}%
  \BibitemOpen
  \bibfield  {author} {\bibinfo {author} {\bibfnamefont {A.~D.}\ \bibnamefont
  {Bandrauk}}\ and\ \bibinfo {author} {\bibfnamefont {H.}~\bibnamefont
  {Shen}},\ }\bibfield  {title} {\bibinfo {title} {Exponential split operator
  methods for solving coupled time‐dependent schrödinger equations},\ }\href
  {https://doi.org/10.1063/1.465362} {\bibfield  {journal} {\bibinfo  {journal}
  {J. Chem. Phys.}\ }\textbf {\bibinfo {volume} {99}},\ \bibinfo {pages} {1185}
  (\bibinfo {year} {1993})}\BibitemShut {NoStop}%
\end{thebibliography}%
\end{document}


\title{Supplementary Materials for ``Quantum Optimal Control without Arbitrary Waveform Generators"}
\author{Qi-Ming Chen}
\affiliation{QCD Labs, QTF Centre of Excellence, Department of Applied Physics, Aalto University, FI-00076, Espoo, Finland}
\affiliation{Department of Chemistry, Princeton University, Princeton, New Jersey 08544, USA}
\affiliation{Department of Automation, Tsinghua University, Beijing 100084, China}

\author{Herschel Rabitz}
\email{hrabitz@princeton.edu}
\affiliation{Department of Chemistry, Princeton University, Princeton, New Jersey 08544, USA}

\author{Re-Bing Wu}
\email{rbwu@tsinghua.edu.cn}
\affiliation{Department of Automation, Tsinghua University, Beijing 100084, China}

\date{\today}
\maketitle

\section{Waveform-pulse train correspondance}\label{s:fourier}
\subsection{single-frequency waveform}
To demonstrate the equivalence between a waveform and a pulse train, we start from the simplest case where the former is a sinusoidal function of time, $u_1(t)=\sin(\omega t + \phi)$. Here, the subscript $k=1$ is a special case of the $k$th control field in a general quantum control system. We describe the corresponding pulse train as a sequence of rectangular pulses centered at $(m-{1}/{2})\tau$ for $m=1,\cdots,M$,  
\begin{align}
	s_1(t) = \xi_1 \sum_{m=1}^{M}\left\{ \theta\left[ t - \left(m-\frac{1}{2}\right)\tau + \frac{\tau_{1,m}}{2} \right] - \theta\left[ t - \left(m-\frac{1}{2}\right)\tau - \frac{\tau_{1,m}}{2} \right] \right\}.
\end{align}
Here, $\theta(t)$ is the unit step function, $M=T/\tau$ is the number of pulses in one period $T=2\pi/\omega$, $\xi_1$ is the magnitude for all the pulses, and $|\tau_{1,m}|$ is the width of the $m$th pulse. By utilizing a Fourier series, the above equation can be equivalently written as 
\begin{align}
  s_1(t)=\frac{\xi_1}{T}\sum_{m=1}^{M}\tau_{1,m}
  +\sum_{n\neq0}\Big[\sum_{m=1}^{M}\frac{\xi_1}{n\pi}
  \sin{\Big(\frac{\tau_{1,m}}{\tau}\frac{n\pi}{M}\Big)}
  e^{-in\omega (m-\frac{1}{2})\tau}\Big]e^{in\omega t}. \label{SUM}
\end{align}
Moreover, we define the signed pulse width and magnitude of the $m$th pulse as
\begin{align}
  \tau_{1,m} = \frac{1}{\xi_1}\int_{(m-1)\tau}^{m\tau}dt \sin(\omega t + \phi),\,
  \xi_1 = \max_{0\leq t\leq T}\left|\sin(\omega t + \phi)\right|.
  \label{eq:TP1}
\end{align}
This definition guarantees that the pulse width, $\left|\tau_{1,m}\right|$, is always smaller than or equal to $\tau$, while its $\pm$ sign determines the polarization of the pulse. 

By inserting Eq.\,\eqref{TP1} into \eqref{SUM}, we can readily eliminate the $1$st term in the Fourier series which is the DC component. For the $2$nd term, we use the following approximation to simplify the summation
\begin{align}
  \sum_{n \ll M}\Big[\sum_{m=1}^{M}\frac{\xi_1}{n\pi} \sin{\Big(\frac{\tau_{1,m}}{\tau}\cdot\frac{n\pi}{M}\Big)}
  e^{-in\omega (m-\frac{1}{2})\tau}\Big]e^{in\omega t}
  \approx\sum_{n \ll M}\Big[\sum_{m=1}^{M}\frac{\xi_1 \tau_{1,m}}{M\tau}e^{-in\omega (m-\frac{1}{2})\tau}\Big]e^{in\omega t}. \label{APPRO}
\end{align}
In general, this expression holds for $n \ll M$. However, it is reasonably precise for all values of $n$ if we assume $M \gg 1$. This conclusion arises because the coefficient $1/n$ becomes sufficiently small in this case and plays a negligibly small contribution to the summation. In this regard, we simplify Eq.\,\eqref{SUM} as
\begin{align}
  s_1(t)= \sum_{n\neq0}\Big[\sum_{m=1}^{M}\frac{\xi_1 \tau_{1,m}}{M\tau}e^{-in\omega (m-\frac{1}{2})\tau}\Big]e^{in\omega t}. \label{SUM2}
\end{align}

For $n=1$ in the summation, we have
\begin{align}
  \frac{\xi_1}{M}\sum_{m=1}^{M}\frac{\tau_{1,m}}{\tau} e^{-i\omega(m-\frac{1}{2})\tau}e^{i\omega t}
  =\frac{e^{i\frac{\pi}{M}}}{4\pi}\Big[
  M\Big(1-e^{i\frac{2\pi}{M}}\Big)e^{i\phi}
  +\Big(1-e^{-i\frac{2\pi}{M}}\Big)e^{-i\phi}\sum_{m=1}^{M}e^{-i\frac{2m}{M}2\pi}\Big]
  e^{i\omega t}
  \approx\frac{e^{i(\omega t+\phi)}}{2i}. \label{eq:nw}
\end{align}
Similarly, for $n=-1$ we have
\begin{align}
  \frac{\xi_1}{M}\sum_{m=1}^{M}\frac{\tau_{1,m}}{\tau} e^{i\omega(m-\frac{1}{2})\tau}e^{-i\omega t}
  =\frac{e^{-i\frac{\pi}{M}}}{4\pi}\Big[
  \Big(1-e^{i\frac{2\pi}{M}}\Big)e^{i\phi}\sum_{m=1}^{M}e^{i\frac{2m}{M}2\pi}
  +M\Big(1-e^{-i\frac{2\pi}{M}}\Big)e^{-i\phi}\Big]e^{-i\omega t}
  \approx-\frac{e^{i(\omega t-\phi)}}{2i}. \label{eq:-nw}
\end{align}
One can verify that all the other terms, from $n=\pm 2$ to $\pm (M-2)$, are equal to \textit{zero}. This observation shows that the pulse sequence $s_1(t)$ can be regarded as a combination of $u_1(t)$ and a high-frequency error term. The frequency threshold is $\Omega_1=(M-1)\omega \approx M\omega$, below which the Fourier components of the pulse sequence, $s_1(t)$, and the waveform, $u_1(t)$, are indistinguishable from each other. The error exists only in high frequency components, which can be reasonably omitted by the same argument behind the rotating wave approximation, or physically filtered before the sample input in experiments. We note that the value of $\Omega_1$ is fully determined by the pulse number $M$, which indicates the ability to adjust the scale of the approximation error for different purposes.

\subsection{Arbitrary waveform}
We now study the general case where $u_1(t)$ is an arbitrary real function. We assume that the waveform has a finite frequency bandwidth $[\omega_{\rm min},\omega_{\rm max}]$, and write it as
\begin{eqnarray}\label{FU}
  u_1(t)
  =2\int_{\omega_{min}}^{\omega_{max}}d\omega
  \left| U_1(\omega) \right| \sin{(\omega t+\phi(\omega))}.
\end{eqnarray}
where $U_1(\omega)$ is the Fourier transform of $u_1(t)$, and $\phi(\omega)=\arctan \left( -{\rm Re}[U_1(\omega)]/{\rm Im}[U_1(\omega)] \right)$. Here, we have used the property that $U_1(-\omega)=U_1^*(\omega)$ for the real function $u_1(t)$. 

Following the developments in the single-frequency case, for each frequency component, $\omega$, one can generate a sequence of PWM pulses to approximate the sinusoidal component $\sin{(\omega t+\phi(\omega))}$. In each short time interval, the signed pulse width, $\tau_{1,m}(\omega)$, is defined in the same way of Eq.\,\eqref{eq:TP1}. We further require that the signed pulse widths for different frequencies must be the same, which is achieved by adjusting the pulse magnitudes, $\xi_{1}(\omega)$, for different frequencies. Assuming that the desired value of the signed pulse width is $\tau_{1,m}$, the pulse magnitude corresponding to the frequency $\omega$ should be calculated as
\begin{align}
	\xi_1(\omega) = \frac{2}{\tau_{1,m}}\int_{(m-1)\tau}^{m\tau}dt U_1(\omega)\sin{(\omega t+\phi(\omega))}.
\end{align}
The pulse magnitude $\xi_1$ in the $m$th time interval is an integral of $\xi_1(\omega)$ over the entire frequency range $[\omega_{\rm min},\omega_{\rm max}]$,
\begin{align}
	\xi_1
	= \frac{2}{\tau_{1,m}}\int_{(m-1)\tau}^{m\tau}dt \int_{\omega_{min}}^{\omega_{max}}d\omega U_1(\omega)\sin{(\omega t+\phi(\omega))}
	= \frac{1}{\tau_{1,m}}\int_{(m-1)\tau}^{m\tau}dt u_1(t).
\end{align}
From a different perspective, if we require the pulse magnitude to be $\xi_1$, the corresponding signed pulse width is 
\begin{align}
  \tau_{1,m} = \frac{1}{\xi_1}\int_{(m-1)\tau}^{m\tau}dt u_1(t).
  \label{TP1}
\end{align}
This means that an arbitrary function $u_1(t)$ with finite frequency range can be also approximated with a pulse train by neglecting high-frequency errors above $\Omega_{1}$. One can prove that the threshold frequency in this case is $\Omega_{1}\approx M\omega_{\rm min}$. So long as $\Omega_{1} \gg \omega_{\rm max}$, or equivalently $\tau \ll 2\pi/\omega_{\rm max}$, the pulse sequence $s_1(t)$ is indistinguishable from $u_1(t)$ below $\Omega_{1}$.

\section{Error analysis}\label{s:time_propagation}
We consider a general quantum system described by the following Hamiltonian
\begin{align}
	H = H_0 + \sum_{k=1}^{K}u_k(t)H_k, \label{eq:original_hamiltonian}
\end{align}
where $H_k$ is Hermitian and $u_k(t)$ is a real function of time. The system with pulse width modulation (PWM) control can be described as
\begin{align}
	H = H_0 + \sum_{k=1}^{K}s_k(t)H_k. \label{eq:pwm_hamiltonian}
\end{align}
We now compare the time propagation of the system driven by the waveforms, $u_k(t)$, and by the pulse trains, $s_k(t)$. 

\subsection{Single-control case}
We start from the simplest case with $K=1$. The time propagator with waveform $u_1(t)$ in the $m$th time interval can be written in a Dyson series as
\begin{align}\label{err}
  U_{u}\left[m\tau,(m-1)\tau\right]
  &=1 -i \sum_{k_1=0}^{1}H_{k_1}\int_{(m-1)\tau}^{m\tau}dtu_{k_1}(t) \nonumber \\
  &+(-i)^2\sum_{k_1,k_2=0}^{1} \left[ H_{k_1} H_{k_2} \int_{(m-1)\tau}^{m\tau}dt_1{\int_{(m-1)\tau}^{t_1}dt_2{u_{k_1}(t_1)u_{k_2}(t_2)}}\right] + O(\tau^3), 
\end{align}
where we have defined $u_0(t)\equiv 1$ to simplify the notation. On the other hand, the corresponding pulsed time propagator can be written as
\begin{align}
  U_{s}\left[m\tau,(m-1)\tau\right]=\exp{\Big(-i |\tau_{0,m}| H_0\Big)}
  \exp{\Big(-i |\tau_{1,m}| \left( H_0 + {\rm sgn}[\tau_{1,m}]\xi_1 H_1 \right) \Big)}
  \exp{\Big(-i |\tau_{0,m}| H_0\Big)}, \label{eq:pwmpropagation}
\end{align}
where $\tau_{1,m} = \int_{(m-1)\tau}^{m\tau}dt u_1(t)/\xi_1$, $\tau_{0,m} = \left( \tau - |\tau_{1,m}| \right)/2$. Because the generators of time propagation are time independent, one can simply use a Taylor series to rewrite the propagator as 
\begin{align}
  U_{s}\left[m\tau,(m-1)\tau\right]
  &=\Big[1-i|\tau_{0,m}| H_0 +\frac{(-i)^2}{2!}\Big(|\tau_{0,m}| H_0\Big)^2+O(\tau^3)\Big]\nonumber\\ 
  &\times\Big[1-i |\tau_{1,m}| \left( H_0 + {\rm sgn}[\tau_{1,m}]\xi_1 H_1 \right) +\frac{(-i)^2}{2!}\Big(|\tau_{1,m}| \left( H_0 + {\rm sgn}[\tau_{1,m}]\xi_1 H_1 \right)\Big)^2+O(\tau^3)\Big] \nonumber\\ 
  &\times\Big[1 -i|\tau_{0,m}| H_0 +\frac{(-i)^2}{2!}\Big(|\tau_{0,m}| H_0\Big)^2+O(\tau^3)\Big] \nonumber\\
  &=1-i\left(\tau H_0 + \xi_{1}\tau_{1,m} H_{1} \right)
  +\frac{(-i)^2}{2!}\left( \tau H_0 + \xi_{1}\tau_{1,m} H_{1} \right)^2+ O(\tau^3).
\end{align}

We define the propagation error as the difference between the two evolution operators 
\begin{align}
  \delta U[m\tau, (m-1)\tau] 
  &= U_{u}\left[m\tau,(m-1)\tau\right]-U_{s}\left[m\tau,(m-1)\tau\right] \nonumber \\
  &= {(-i)^2}\sum_{k_1,k_2=0}^{1} H_{k_1} H_{k_2} \Big(\int_{(m-1)\tau}^{m\tau}dt_1{\int_{(m-1)\tau}^{t_1}dt_2{u_{k_1}(t_1)u_{k_2}(t_2)}} \nonumber \\
  &-\frac{1}{2!}\int_{(m-1)\tau}^{m\tau}dt_1 u_{k_1}(t_1) {\int_{(m-1)\tau}^{m\tau}dt_2{u_{k_2}(t_2)}}\Big)
  + O(\tau^3). \label{eq:err}
\end{align}
To simplify the time-ordered integral in the above equation, we use the following relations  
\begin{align}
  &\int_{(m-1)\tau}^{m\tau}dt{u_{k}(t)}
  =u_{k}(t')\tau+\frac{1}{2!}u'_{k}(t')\tau^2+O(\tau^3), \label{eq:tint2} \\
  &\int_{(m-1)\tau}^{m\tau}dt_1{\int_{(m-1)\tau}^{t_1}dt_2{u_{k_1}(t_1)u_{k_2}(t_2)}}
  =\frac{1}{2}u_{k_1}(t')u_{k_2}(t')\tau^2+O(\tau^3), \label{eq:tint}
\end{align}
Here, $t'$ is an arbitrary time instance within $[(m-1)\tau, m\tau]$. By inserting Eqs.\,\eqref{eq:tint} and\eqref{eq:tint2} into \eqref{eq:err}, we obtain $\delta U[m\tau, (m-1)\tau] = O(\tau^3)$. This result shows that the pulse propagation has $2$nd-order accuracy in simulating the dynamics driven by a continuous waveform. 

\subsection{Multi-control case}
For a general quantum system with $K$ controls, the pulse time propagation is 
\begin{align}
  U_{s}\left[m\tau,(m-1)\tau\right]
  &=\exp{\left(-i |\tau_{0,m}| H_0\right)}
  \exp{\left[-i |\tau_{1,m}| \left( H_{0} + {\rm sgn}[\tau_{1,m}]\xi_{1} H_{1} \right) \right]} \cdots \nonumber\\
  &\times \exp{\left[-i |\tau_{K,m}| \left( H_{0} + \sum_{k=1}^{K}{\rm sgn}[\tau_{k,m}]\xi_{k} H_{k} \right) \right]}
  \exp{\left[-i |\tau_{(K-1),m}| \left( H_0 + \sum_{k=1}^{K-1}{\rm sgn}[\tau_{k,m}]\xi_{k} H_{k} \right) \right]}
  \cdots \nonumber\\
  &\times \exp{\left(-i |\tau_{0,m}| H_0\right)},
\end{align}
where
\begin{align}
	|\tau_{k,m}| = \begin{cases}
		\frac{1}{2}\left[\tau - \frac{1}{\xi_{1}}\left|\int_{(m-1)\tau}^{m\tau}dt u_{1}(t)\right|\right],\,&\text{for}\,k=0; \\
		\frac{1}{2}\left[\frac{1}{\xi_{k}}\left|\int_{(m-1)\tau}^{m\tau}dt u_{k}(t)\right|
	- \frac{1}{\xi_{k+1}}\left|\int_{(m-1)\tau}^{m\tau}dt u_{(k+1)}(t)\right|\right],\,&\text{for}\,1 \leq k \leq K-1; \\
		\frac{1}{\xi_{K}}\left|\int_{(m-1)\tau}^{m\tau}dt u_{K}(t)\right|,\,&\text{for}\,k=K.
	\end{cases}
\end{align}
Here, to simplify the notation we have relabelled the control fields such that $|\tau| > |\tau_{1,m}|>\cdots>|\tau_{K,m}|$. We note that the $\pm$ sign of $\tau_{k,m}$ is determined by the $\pm$ sign of the integral $\int_{(m-1)\tau}^{m\tau}dt u_{k}(t)$ only. The physical meaning is that we align all the pulses for different $k$ to the same center of each time interval. Thus, the interaction ${\rm sgn}[\tau_{k,m}]\xi_{k}H_{k}$ with larger pulse area is turned on earlier and turned off later compared with the others. Following the same procedure in the single control case, we use the Taylor series to rewrite each exponential to $2$nd order and recombine the terms according to $\tau$. One can prove that the pulse propagation is still $2$nd-order accurate for $K > 1$. This result is as precise as the staircase approximation of an arbitrary function, which is widely used in numerical simulations and is utilized by an arbitrary waveform generator. 

\section{Numerical efficiency} \label{s:speed_up}
In numerical simulation, the replacement, $u_k(t) \rightarrow s_k(t)$, has an advantage that the resulting generator of time propagation belongs to a finite set $H(t) \in \left\{H_0, H_0 + \xi_1 H_1, H_0 - \xi_1 H_1,\cdots,H_0 + \cdots + \xi_K H_K, \cdots, H_0 - \cdots - \xi_K H_K \right\}$. This observation indicates that the numerically expensive calculation of matrix exponents in conventional design efforts can be circumvented with diagonal matrix exponents instead. In detail, the pulse time propagation can be written as
\begin{align}
  U_{s}\left[m\tau,(m-1)\tau\right]
  &=P_0\exp\left[-i\left(\tau-\frac{|\tau_{1,m}|}{2}\right)\Lambda_0\right] P_0^{\dagger}
	P_{1} \exp\left[-i|\tau_{1,m}|\Lambda_{1}\right] P_{1}^{\dagger} \cdots 
	P_{K} \exp\left[-i|\tau_{K,m}|\Lambda_{K}\right] P_{K}^{\dagger}\nonumber \\
	&\times P_{K-1} \exp\left[-i|\tau_{K-1,m}|\Lambda_{K-1}\right] P_{K-1}^{\dagger} \cdots
	P_0 \exp\left[-i\left(\tau-\frac{|\tau_{1,m}|}{2}\right)\Lambda_0\right] P_0^{\dagger}.
\end{align}
Here, $\Lambda_{k}$ and $P_{k}$ are real diagonal and unitary matrices, respectively. For $k=0$, it is defined as $H_0 = P_{0}\Lambda_{0}P_{0}^{\dagger}$. However, the calculation of $\Lambda_{k}$ for $k=1,2,\cdots,k$ depends on the polarization of $s_k(t)$. For example, if the integrals of the control fields are $\tau_{1,m}<0$, $\tau_{2,m}<0$, and $\tau_{3,m}\geq 0$ in the $m$th interval, we have $H_0 - \xi_1 H_1 = P_{1}\Lambda_{1}P_{1}^{\dagger}$, $H_0 - \xi_1 H_1 - \xi_2 H_2 = P_{2}\Lambda_{2}P_{2}^{\dagger}$, $H_0 - \xi_1 H_1 - \xi_2 H_2 + \xi_3 H_3 = P_{3}\Lambda_{3}P_{3}^{\dagger}$. Thus, one has to determine the polarization of $s_k(m)$ before calculating the propagator. Because the set is finite, one can diagonalize different terms in advance and save the result. This procedure is particularly efficient in Python by using its dictionary function. In comparison, the time propagation with the control field, $u_1(t)$, is 
\begin{align}
	U_{u}(T,0) = \prod_{m=1}^{M}\exp\left[ -i \tau\left(H_0 + \sum_{k=1}^{K}u_k(m\tau)H_k \right) \right]. \label{eq:propagateOri}
\end{align}
Here, the value of $u_k(t)$ can be arbitrary such that the generator of time propagation consists of a convex set $\{H_0+\sum_k u_k(t) H_k| t\in[0,T]\}$. One cannot simply diagonalize the generators, and thus matrix exponentials are unavoidable. 

In Fig.~2(c) of the main text, we compare the numerical efficiency of the two approaches in simulating a $d$-dimensional system. The calculation is carried out under Python 3.7.3 on Windows 10 64-bit (3.20 GHz Intel Core i7-8700, 16 GB). The computational time is measured by using the \textit{perf\_counter()} command in the \textit{time} package, and the matrix product is calculated by the \textit{@} operator. The default method for calculating the matrix exponential refers to the \textit{linalg.expm()} command in the \textit{scipy} package, which is based on the scaling and squaring method \cite{Moler2003}. We observe that a Matlab implementation of the PWM method seems to be less efficient than Python, which we attribute to the possibly lower efficiency of its dictionary function. In the future, even higher numerical advantages may be achieved by using JIT (just in time compiling), cython (C extension in Python), cuda (GPU proceessing), and TensorFlow. 

\section{Generalizations}
\subsection{$n$-level PWM}
In Sections\,\ref{s:fourier}-\ref{s:time_propagation}, we have defined the pulse magnitude of the pulse train as a constant, $\xi_k$. The polarization of the pulse is determined by the short-time integral of $u_k(t)$, i.e., the sign of $\tau_{k,m}$ defined in Eq.\,\eqref{TP1}. This is called the 3-level PWM in the literature of power electronics as the field at any time instance is chosen from three candidate values \cite{Holmes2003}. Besides, there are many other forms of PWM, among which the 2-level PWM is widely used in classical control systems \cite{Holmes2003}. Different from the 3-level case, the 2-level PWM approximates an arbitrary function by switching between two states, as shown in Fig.\,\ref{fig_shaped} (a). If we define the two possible magnitudes as $\xi_{k}$ and $\xi_{k'}$, the time durations in each state, $\tau_{k,m}$ and $\tau_{k',m}$, satisfy the following relation
\begin{align}
	\xi_{k}\tau_{k,m} + \xi_{k'}\tau_{k',m} = \int_{(m-1)\tau}^{m\tau}dt u_k(t),
\end{align}
where $\tau_{k,m}+\tau_{k',m}=\tau$. One can follow the same procedure in Sections\,\ref{s:fourier}-\ref{s:time_propagation}, and prove that the 2-level PWM pulse sequences can also achieve full control of a quantum system.

\subsection{Choice of magnitude} \label{subs:ssp}
In Sections\,\ref{s:fourier}-\ref{s:time_propagation}, we have defined the pulse magnitude as $\xi_k=\max |u_k(t)|,~t\in[0,T]$, which is the maximum value of the waveform over the entire control time. However, the magnitude can also be larger. The extreme case is achieved when the pulse magnitude approaches to the infinity $\xi_k\rightarrow\infty$, indicating the so-called hard pulse. Correspondingly, the pulse width should be shrunk by the same factor to fulfill the requirement of Eq.\,\eqref{TP1}. This extreme case describes a sequence of instantaneous kicks to the system, and the corresponding evolution operator is (assuming that all the pulses are positive)
\begin{eqnarray}
  U_{s}\left[m\tau,(m-1)\tau\right]&=&\exp{\left(-i\tau H_0/2\right)}
  \exp{\left(-i\tau_{1,m} \xi_1 H_1/2\right)}
  \cdots \exp{\left(-i\tau_{k,m} \xi_K H_K/2\right)} \nonumber \\
  &&\times \exp{\left(-i\tau_{k,m} \xi_K H_K/2\right)}
  \cdots \exp{\left(-i\tau_{1,m} \xi_1 H_1/2\right)}
  \exp{\left(-i\tau H_0/2\right)}. \label{eq:spo}
\end{eqnarray}
The advantage of choosing an arbitrarily large magnitude is that the number of generators of time propagation is reduced from $(2^K-K)$ to $(K+1)$, which simplifies the numerical simulation. However, as can be seen from the Fourier analysis in Eq.\,\eqref{SUM2}, the scale of the high-frequency errors depends on the pulse magnitude. Thus, the simulation accuracy will be slightly lower than with the default PWM sequence. In addition, a physical implementation of such intensive pulses may lead to less favorable control fidelities owing to the power limit of real devices.

\subsection{Gaussian pulses}
\begin{figure}[hbt]
  \centering
  \includegraphics[width=0.45\columnwidth]{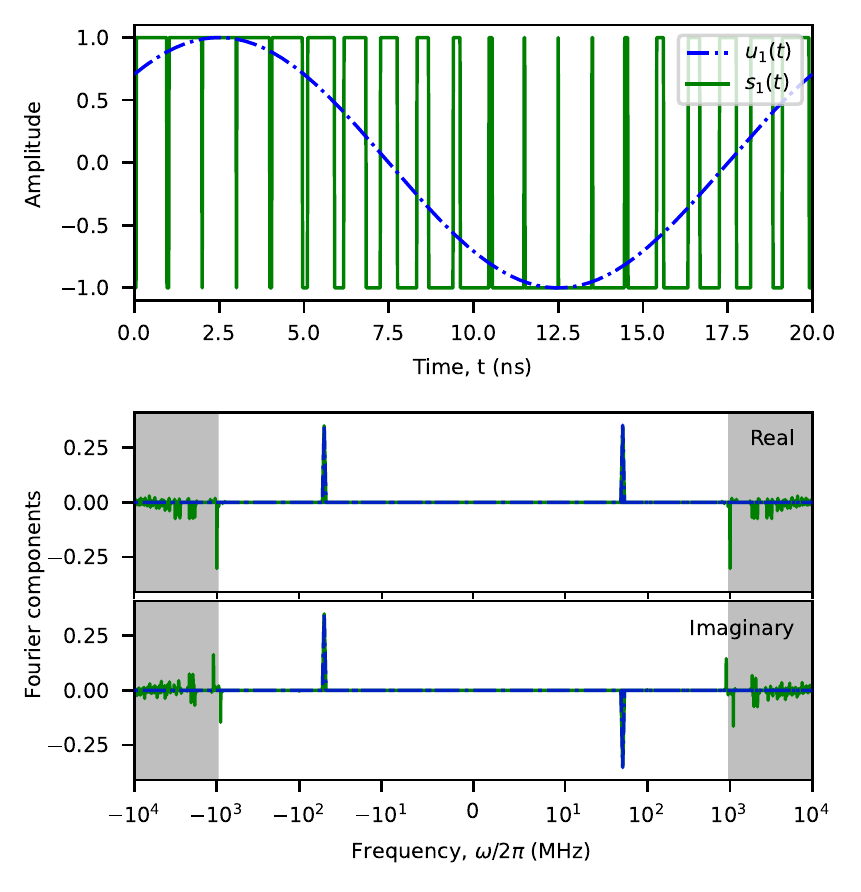}
  \includegraphics[width=0.45\columnwidth]{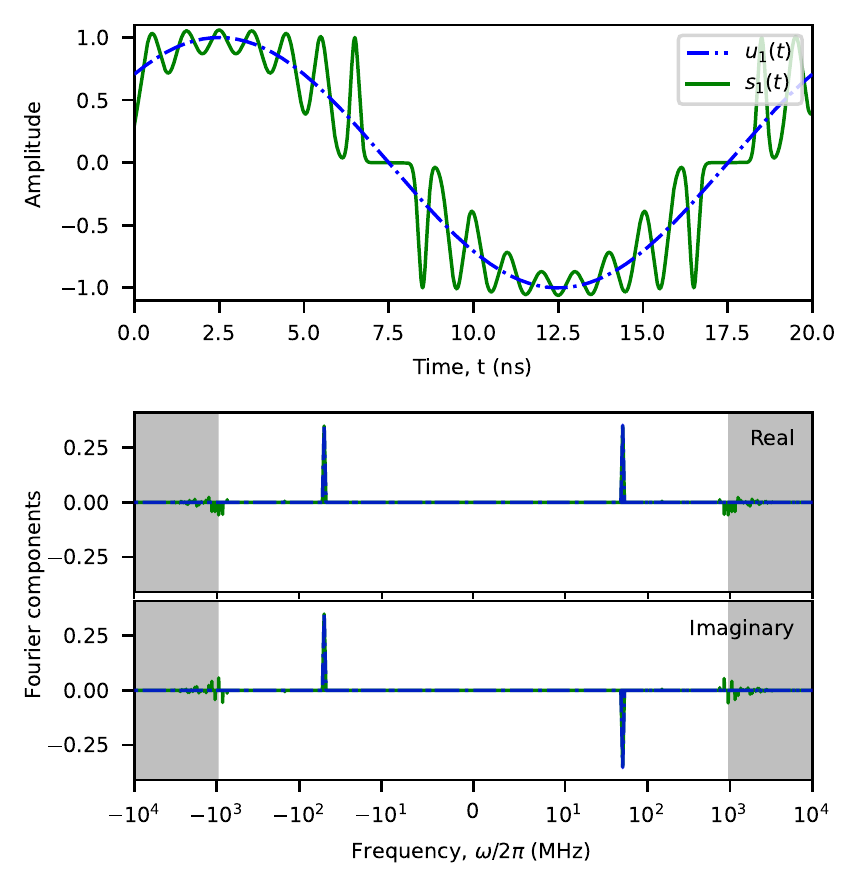}
  \caption{Two-level and Gaussian PWM sequences (left, right). The frequency threshold $\Omega_{1}\approx M \omega_{\rm min}$ is set as $M=20$ times the minimum frequency of $u_1(t)$, i.e., $\Omega_{1}/2\pi \approx 1\,{\rm GHz}$. The frequency components between the maximum frequency of $u_1(t)$ and $\Omega_{1}$ are, in principle, entirely suppressed by properly designed PWM pulses with $\Omega_{1} \gg \omega_{\rm max}$. The difference between $u_1(t)$ and $s_1(t)$ exists only above $\Omega_{1}$ (grey area).}
  \label{fig_shaped}
\end{figure}

In Sections\,\ref{s:fourier}-\ref{s:time_propagation}, we have defined the pulses to be perfect rectangular. However, in real implementations one may expect to use a sequence of smooth pulses instead, considering the power and bandwidth specifications of real control protocols. Here, we consider a sequence of Gaussian pulses, as shown in Fig.~\ref{fig_shaped}(b), which is described as
\begin{align}
	s_k(t)= \xi_k \sum_{m=1}^{M} e^{-\pi\left(t-(m-1/2)\tau \right)/\tau_{k,m}}.
\end{align}
It can also be rewritten as follows by using the Fourier series
\begin{align}\label{SUMG}
  s_k(t)=\sum_{n=-\infty}^{+\infty}\Big(\frac{\xi_k}{T}
  \sum_{m=1}^{M}\tau_{k,m} e^{-\frac{(n\omega t^{(m)})^2}{4\pi}}e^{-in\omega (m-\frac{1}{2})\tau}\Big)e^{in\omega t}.
\end{align}
Following the same analyses in Sec.~\ref{s:fourier}, one can prove that the Fourier components of $u_k(t)$ and $s_k(t)$ are exactly the same below the threshold $\Omega_{k}\approx M\omega_{\rm min}$, as long as
\begin{align}
	\tau_{k,m}=\frac{1}{\xi_k}\int_{(m-1)\tau}^{m\tau}dt u_k(t).
\end{align}
This is exactly the same definition of the signed pulse width for default rectangular PWM pulses, i.e., Eq.\,\eqref{TP1}. 

\subsection{Higher-order expansion}
In Sections\,\ref{s:fourier}-\ref{s:time_propagation}, we proved that the PWM pulse sequence, $s_k(t)$, is a $2$nd-order accurate approximation for simulating the time-propagation driven by $u_k(t)$. This result is based on the assumption that there exists only one pulse in each time interval. To achieve higher-order accuracy in time propagation, one can follow the same procedure introduced in Refs.~\onlinecite{Bandrauk1991, Bandrauk1992, Bandrauk1993}, and insert more pulses in each time interval. We write the exact short-time propagation of $u_k(t)$ as
\begin{align}\label{sub1}
  U_{u}[m\tau, (m-1)\tau]=S_2^{(m)}(\tau)+C_3^{(m)}\tau^3+O(\tau^4),
\end{align}
where $S_2^{(m)}(\tau)=U_{s}[m\tau, (m-1)\tau]$ is the time-propagation driven by PWM pulses, i.e., Eq.\,\eqref{eq:pwmpropagation}. On the other hand, the exact evolution can also be written as
\begin{align}\label{sub2}
  U_{u}(m\tau, (m-1)\tau)=U_{\{u_k\}}(m\tau, (m-s)\tau)U_{\{u_k\}}((m-s)\tau, (m-1+s)\tau)U_{\{u_k\}}((m-1+s)\tau, (m-1)\tau).
\end{align}
Combining Eq.\,\eqref{sub1} and \eqref{sub2}, we have
\begin{eqnarray}
  U_{u}(m\tau, (m-1)\tau)&=&\Big[S_2^{(m)}(s\tau)+ C_3^{(m)}s^3\tau^3+ O(\tau^4)\Big]
  \Big[S_2^{(m)}[(1-2s)\tau]+ C_3^{(m)}(1-2s)^3\tau^3+ O(\tau^4)\Big] \nonumber \\
  &&\times \Big[S_2^{(m)}(s\tau)+ C_3^{(m)}s^3\tau^3+ O(\tau^4)\Big] \nonumber \\
  &=&S_2^{(m)}(s\tau)S_2^{(m)}[(1-2s)\tau]S_2^{(m)}(s\tau)+ C_3^{(m)}\tau^3[2s^3+(1-2s)^3]+ O(\tau^4).
\end{eqnarray}
When $2s^3+(1-2s)^3=0$, we obtain the PWM sequence with $3$rd-order accuracy 
\begin{align}
  U_{s}\left[m\tau,(m-1)\tau\right]=S_3^{(m)}(\tau)=U_{\{u_1\},m}+ O(\tau^4),
\end{align}
where
\begin{align}\label{high}
  S_3^{(m)}(\tau)=S_2^{(m)}(s\tau)S_2^{(m)}[(1-2s)\tau]S_2^{(m)}(s\tau).
\end{align}
Similarly, one can obtain an arbitrarily high-order form of PWM. We note that although the solution for $s$ always exists mathematically, the value of $s$ may not be positive and real in higher-order expansions. The propagator, $S_{n}^{(m)}(\tau)$, in such cases cannot be physically implemented because the polarization of the drift Hamiltonian, $H_0$, is not controllable. However, one can prove that there is always a real and positive solution $s=1/(2-2^{1/(2n+1)})$ for the $(2n+1)$th-order form of PWM, $S_{2n+1}^{(m)}(\tau)$ \cite{Bandrauk1991, Bandrauk1992, Bandrauk1993}.

\bibliography{PWM_REF}